\documentclass[showpacs,preprintnumbers,amsmath,amssymb,nofootinbib,superscriptaddress,floatfix]{revtex4}
\usepackage[normalem]{ulem}
\usepackage{graphicx}
\usepackage{dcolumn}
\usepackage{bm}
\usepackage{color}
\usepackage[colorlinks,citecolor=blue,urlcolor=blue,linkcolor=blue]{hyperref}

\usepackage{subfigure}

\newcommand{\f}{\frac}
\newcommand{\lt}{\left}
\newcommand{\n}{\nonumber}
\newcommand{\p}{\partial}
\newcommand{\rd}{{\rm d}}
\newcommand{\rt}{\right}
\newcommand{\ve}{\varepsilon}

\newcommand{\arXg}[1]{\href{http://arxiv.org/abs/#1}{{\ttfamily arXiv:#1[gr-qc]}}}
\newcommand{\arXh}[1]{\href{http://arxiv.org/abs/#1}{{\ttfamily arXiv:#1[hep-th]}}}

\begin{document}
\title{On the throttling process of the Kerr--Newman--anti-de Sitter black holes in the extended phase space}

\author{Ze-Wei Zhao}
\affiliation{Department of Physics, College of Sciences, Northeastern University, Shenyang 110819, China}
\author{Yi-Hong Xiu}
\affiliation{Department of Physics, College of Sciences, Northeastern University, Shenyang 110819, China}
\author{Nan Li}
\email{linan@mail.neu.edu.cn. Corresponding author.}
\affiliation{Department of Physics, College of Sciences, Northeastern University, Shenyang 110819, China}
\date{Received: date / Accepted: date}

\begin{abstract}
The throttling process of the Kerr--Newman--anti-de Sitter (KN--AdS) black holes is systematically studied in the extended phase space. In this framework, the cosmological constant is interpreted as a varying thermodynamic pressure, and the black hole mass is identified with enthalpy. The throttling process is essentially an isenthalpic (i.e., constant-mass) process for the KN--AdS black holes. The Joule--Thomson coefficient, inversion temperature, inversion curve, and isenthalpic curve are investigated in order, with both analytical and numerical methods. It is found that there are no maximum inversion temperatures, but only minimum ones that are around one half of the critical temperatures of the KN--AdS black holes. Two characteristic masses are also discussed to show the detailed features in the throttling behaviors of the KN--AdS black holes.
\end{abstract}

\maketitle

\section{Introduction} \label{sec:intro}

Black hole thermodynamics has been one of the most exciting topics in modern theoretical physics during the past half century, as it shows the miraculous relationship among thermodynamics, classical gravity, and quantum mechanics, and thus provides a promising path to our final understanding of quantum gravity. The research was pioneered by the introduction of black hole entropy \cite{Bekenstein}, succeeded by the establishment of black hole thermodynamic laws \cite{law}, and eventually integrated in the discovery of the Hawking radiation \cite{Hawking2}. These studies indicate that a black hole is not merely a simple mechanical object, but should be considered as a complicated thermodynamic system with temperature and entropy.

A black hole may exhibit rich thermodynamic behaviors, especially in the anti-de Sitter (AdS) space-time with a negative cosmological constant. The thermodynamics of the AdS black holes was originally studied in the Hawking--Page phase transition between a stable Schwarzschild--AdS black hole and the thermal gas in the AdS space \cite{HP}. This work was extended to the charged AdS black holes [i.e., Reissner--Nordstr\"{o}m--AdS (RN--AdS) black holes] in Refs. \cite{Chamblin1, Chamblin2, B1, B2}, and an analogy was found between the RN--AdS black holes and the van der Waals fluids in their phase diagrams.

Recently, the attempt to consider the cosmological constant as a varying thermodynamic pressure, rather than a fixed AdS background, has attracted increasing attention in the literature \cite{Cald, Kastor, Kastor1, Dolan1, Dolan2, Cvetic}. Allowing the variation of the cosmological constant is equivalent to adding a new dimension in the black hole thermodynamic phase space, so such a theory is named as ``black hole thermodynamics in the extended phase space''. In Ref. \cite{KM}, this idea together with the critical phenomena of the RN--AdS black holes was systematically studied by Kubiz\v{n}\'{a}k and Mann, in which the black hole mass is identified as enthalpy rather than internal energy, and the cosmological constant $\Lambda$ plays the role of the thermodynamic pressure as
\begin{align}
p=-\f{\Lambda}{8\pi}=\f{3}{8\pi l^2}, \label{P1}
\end{align}
where $l$ is the AdS curvature radius, and the conjugate variable of $p$ can be defined as the black hole thermodynamic volume, even if the geometric volume cannot be defined inside the horizon. By this means, it is direct to find the correspondence between the RN--AdS black holes and the van der Waals fluids, such as the similarities in the small--large black hole phase transitions and the gas--liquid phase transitions. These remarkable observations aroused a large number of successive studies. It was soon found that not only the RN--AdS black holes, but other various black holes also possess the van der Waals-like behaviors. Moreover, the thermodynamic geometry \cite{geometry}, compressibility \cite{compre}, heat cycle \cite{cycle}, Maxwell construction \cite{wsw}, Ehrenfest scheme \cite{Eh}, triple point \cite{triple}, reentrant phase transition \cite{trans}, and critical phenomenon \cite{crip} of these black holes, especially in the modified gravity theories, were widely investigated. The recent progresses can be found in Refs. \cite{rev1, rev3} for excellent reviews.

Among these studies, an interesting thermodynamic issue, the throttling process [i.e., Joule--Thomson (JT) effect], has received more and more interest. In Ref. \cite{okcu1}, \"{O}kc\"{u} and Ayd{\i}ner explored this process for the RN--AdS black holes for the first time and obtained the inversion curves and isenthalpic curves. These authors also studied the rotating AdS black holes (i.e., Kerr--AdS black holes) in Ref. \cite{okcu2} and found similar results. Some relevant research can also be found in Refs. \cite{D'Almeida,Gh, Mo, Chabab, Mo2018, Lan} for the holographic superfluids, for the quintessential RN--AdS black holes, and for the RN--AdS black holes in high-dimensional space-times, in $f(R)$ gravity, in Lovelock gravity, and in Gauss--Bonnet gravity. All these studies further confirmed the basic conclusions in Ref. \cite{okcu1}.

Albeit the throttling processes of various black holes have been intensively studied, there still leaves an very important case not yet investigated---the four-dimensional rotating and charged AdS black holes [i.e., Kerr--Newman--AdS (KN--AdS) black holes]. It is quite natural to proceed to this most general black hole solution immediately after the special RN--AdS and Kerr--AdS black holes, rather than to investigate the complicated black holes in various modified gravity theories. However, as we will show in Sect. \ref{sec:inv}, a mathematical obstacle prevented the direct generalization. Consequently, most previous studies concentrated on the AdS-like black holes with simple spherical horizon topology \cite{okcu1, D'Almeida, Gh, Mo, Chabab, Mo2018, Lan}. Otherwise, the inversion curves could hardly be attainable. In the present work, we will show how to solve this problem and then thoroughly study the throttling process of the KN--AdS black holes. Several relevant issues will also be carefully discussed, as some of them have not been considered in detail before. Altogether, we wish to give a complete picture of the throttling process of the KN--AdS black holes in the extended phase space to the most general extent.

This paper is organized as follows. In Sect. \ref{sec:KNAdS}, we briefly list the thermodynamic properties of the KN--AdS black holes in the extended phase space. In Sect. \ref{sec:throttling}, we first explain the throttling process of the KN--AdS black holes and then investigate the JT coefficient, inversion temperature, inversion curve, and isenthalpic curve in Sects. \ref{sec:coefficient}--\ref{sec:iso}. The similarities and differences between our results and the previous ones in Refs. \cite{okcu1, okcu2, D'Almeida, Gh, Mo, Chabab, Mo2018, Lan} are also presented at the same time. We conclude in Sect. \ref{sec:con}. In this paper, we work in the natural system of units and set $c=G_{\rm N}=\hbar=k_{\rm B}=1$.

\section{Thermodynamics of the KN--AdS black holes in the extended phase space} \label{sec:KNAdS}

In this section, we briefly review the necessary thermodynamic properties of the KN--AdS black holes in the extended phase space. The metric of the KN--AdS black hole expressed in the Boyer--Lindquist-like coordinates reads \cite{Carter}
\begin{align}
\rd s^2&=-\f{\Delta_r}{\rho^2}\lt(\rd t-\f{a\sin^2\theta}{\Xi}\,\rd\phi\rt)^2+\f{\rho^2}{\Delta_r}\,\rd r^2+\f{\rho^2}{\Delta_\theta}\,\rd\theta^2+\f{\sin^2\theta\Delta_\theta}{\rho^2}\lt(a\,\rd t-\f{r^2+a^2}{\Xi}\,\rd\phi\rt)^2, \n
\end{align}
where
\begin{align}
\rho^2=r^2+a^2\cos^2\theta, \quad \Xi=1-\f{a^2}{l^2}, \quad \Delta_r=(r^2+a^2)\lt(1+\f{r^2}{l^2}\rt)-2mr+q^2, \quad \Delta_\theta=1-\f{a^2}{l^2}\cos^2\theta, \label{shijie}
\end{align}
and $m$, $q$, and $a$ character the mass $M$, charge $Q$, and angular momentum $J$ of the KN--AdS black hole respectively,
\begin{align}
M=\f{m}{\Xi^2}, \quad Q=\f{q}{\Xi}, \quad J=aM=\f{am}{\Xi^2}. \label{MQJ}
\end{align}

From Eq. (\ref{shijie}), we can fix the horizon radius $r_+$ as the largest root of $\Delta_r=0$ and simultaneously obtain the mass of the KN--AdS black hole as
\begin{align}
M=\f{(r_+^2+a^2)(r_+^2+l^2)+q^2l^2}{2r_+l^2\Xi^2}. \label{zhongjian}
\end{align}
To avoid naked singularity, $r_+$ must be positive, and this sets a constraint on the values of $M$, $J$, and $Q$, namely $2M^2>\sqrt{4 J^2+Q^4}+Q^2$. This inequality reduces to $M>Q$ for the RN--AdS black hole and $M>\sqrt{J}$ for the Kerr--AdS black hole. Moreover, the Bekenstein--Hawking entropy $S$ of the KN--AdS black hole can be obtained by evaluating the horizon area $A$,
\begin{align}
S=\f{A}{4}=\f{\pi(r_+^2+a^2)}{\Xi}. \n
\end{align}
Solving $r_+$ from $S$, substituting it into Eq. (\ref{zhongjian}), and using Eqs. (\ref{P1}) and (\ref{MQJ}), we can reexpress the mass of the KN--AdS black hole in terms of the thermodynamic variables, $S$, $p$, $J$, and $Q$, as
\begin{align}
M=\f12\sqrt{\f{S}{\pi}\lt[\lt(1+\f{\pi Q^2}{S}+\f{8pS}{3}\rt)^2+ \f{4\pi^2J^2}{S^2}\lt(1+\f{8pS}{3}\rt)\rt]}. \label{M}
\end{align}

The differential form of Eq. (\ref{M}) is just the first law of thermodynamics for the KN--AdS black hole in the extended phase space,
\begin{align}
\rd M=T\,\rd S+V\,\rd p+\Omega\,\rd J+\Phi\,\rd Q, \label{first}
\end{align}
where $T$, $V$, $\Omega$, and $\Phi$ are the Hawking temperature, thermodynamic volume, angular velocity, and electric potential respectively. From Eq. (\ref{first}), we clearly observe that the mass $M$ of the KN--AdS black hole should be essentially identified as enthalpy rather than internal energy. Then, it is straightforward to have
\begin{align}
T&=\lt(\f{\p M}{\p S}\rt)_{p,J,Q}=\f{1}{8\pi M}\lt[\lt(1+\f{\pi Q^2}{S}+\f{8pS}{3}\rt)\lt(1-\f{\pi Q^2}{S}+8pS\rt)-\f{4\pi^2J^2}{S^2}\rt], \label{T}\\
V&=\lt(\f{\p M}{\p p}\rt)_{S,J,Q}=\f{2S^2}{3\pi M}\lt(1+\f{\pi Q^2}{S}+\f{8pS}{3}+\f{2\pi^2J^2}{S^2}\rt), \label{V}\\
\Omega&=\lt(\f{\p M}{\p J}\rt)_{S,p,Q}=\f{\pi J}{MS}\lt(1+\f{8pS}{3}\rt), \label{Omega}\\
\Phi&=\lt(\f{\p M}{\p Q}\rt)_{S,p,J}=\f{Q}{2M}\lt(1+\f{\pi Q^2}{S}+\f{8pS}{3}\rt). \label{Phi}
\end{align}
Furthermore, the Smarr relation \cite{Smarr} can be easily obtained by a scaling argument,
\begin{align}
M=2TS-2pV+2\Omega J+\Phi Q. \label{smarr}
\end{align}

Last, in the extended phase space, the leading-order terms of the equation of state of the KN--AdS black hole can be extracted from Eqs. (\ref{P1}), (\ref{T}), and (\ref{V}) as \cite{Gunasekaran}
\begin{align}
p=\frac{T}{v}-\frac{1}{2\pi v^2}+\frac{2Q^2}{\pi v^4}+\frac{48J^2}{\pi v^6}, \label{eos}
\end{align}
where $v=2[3V/(4\pi)]^{1/3}$ as the specific volume is used for convenience. Therefore, the KN--AdS black hole possesses an equation of state like a non-ideal fluid and thus has rich critical phenomena. In Ref. \cite{liu}, the coexistence lines and the critical values of the thermodynamic variables of the KN--AdS black holes (e.g., critical temperatures and critical pressures) were accurately calculated, and we will consult these values for comparison in the next section.

\section{Throttling process of the KN--AdS black holes in the extended phase space} \label{sec:throttling}

In this section, we explore an important thermodynamic process, the throttling process, for the KN--AdS black holes. In conventional thermodynamics, a throttling process is an adiabatic expansion that a non-ideal fluid (gas or liquid) is forced through a valve or porous plug by a pressure difference. During this process, the fluid may experience a temperature change, so the throttling process can usually be used to cool or liquify a gas even at room temperature and thus has widespread applications in thermal engineering. Although the throttling process is inherently irreversible, the enthalpies of the initial and final states of the fluid do remain unchanged, so an isenthalpic process can be applied to calculate the temperature change.

The temperature change in the throttling process is encoded in the JT coefficient,
\begin{align}
\mu=\lt(\f{\p T}{\p p}\rt)_{H}=\f{1}{C_p}\lt[T\lt(\f{\p V}{\p T}\rt)_{p}-V\rt], \label{JT}
\end{align}
where $C_p=T({\p S}/{\p T})_{p}$ is the heat capacity at constant pressure, and the subscript $H$ indicates that the partial derivative is performed in an isenthalpic process and will be replaced by $M$ in the extended phase space. The sign of $\mu$ separates the $T$--$p$ plane into two regions: the cooling region with $\mu>0$ and the heating region with $\mu<0$, as pressure always decreases in the throttling process. Obviously, if $\mu=0$, there is an inversion temperature $T_{\rm i}$ as the solution of
\begin{align}
T=V\lt(\f{\p T}{\p V}\rt)_{p}, \label{Ti1}
\end{align}
and the $T_{\rm i}$--$p$ relation is the inversion curve.

In the extended phase space, the KN--AdS black holes behave like the non-ideal fluids and can thus experience the throttling process. Of course, there is no valve or porous plug in the universe, so what we really mean is just an adiabatic and isenthalpic process of the KN--AdS black holes. This is physically meaningful. The Hawking radiation of a black hole is usually extremely tiny, so it is quite safe to regard the process adiabatic. With radiation ignored, the black hole mass (i.e., enthalpy in the extended phase space) keeps constant, so the process is also isenthalpic. Therefore, both aspects satisfy the necessary conditions in the throttling process. Below, we study the JT coefficient, inversion temperature, inversion curve, and isenthalpic curve of the KN--AdS black holes in order.

\subsection{JT coefficient} \label{sec:coefficient}

We start from the JT coefficient. First, from Eq. (\ref{T}), the heat capacity at constant pressure of the KN--AdS black hole is
\begin{align}
C_p=\f{24\pi M^2S^2\{S^2[3+16\pi pQ^2+32pS(1+2pS)]-3\pi^2(4J^2+Q^4)\}}
{48\pi M^2S[3\pi^2(4J^2+Q^4)+16pS^3(1+4pS)]-\{12\pi^2J^2+[\pi Q^2-S(1+8pS)][3\pi Q^2+S(3+8pS)]\}^2}. \label{Cp1}
\end{align}
Then, using Eqs. (\ref{V}) and (\ref{Cp1}), we obtain the JT coefficient in Eq. (\ref{JT}),
\begin{align}
\mu&=\lt(\f{\p V}{\p S}\rt)_{p,J,Q}-\f{V}{C_p} \n\\
&=\f{288\pi^4J^4+2S[3\pi Q^2-2S(1+4pS)][3\pi Q^2+S(3+8pS)]^2+24\pi^2J^2[3\pi Q^2(\pi Q^2+3S)+4S^2(1+2pS)(3+8pS)]}
{9\pi MS\{12\pi^2J^2+[\pi Q^2-S(1+8pS)][3\pi Q^2+S(3+8pS)]\}}. \label{mu1}
\end{align}
Attention, in Eqs. (\ref{Cp1}) and (\ref{mu1}), $M$ is still a function of $S$, $p$, $J$, and $Q$, as can be seen in Eq. (\ref{M}). Moreover, Eq. (\ref{mu1}) indicates that a divergent $C_p$ cannot guarantee a vanishing $\mu$ as Eq. (\ref{JT}) apparently shows, so the inversion temperatures and inversion curves will be discussed in more detail in Sects. \ref{sec:inv} and \ref{sec:invc}.

\begin{figure}[h]
\begin{center}
\includegraphics[width=0.45\linewidth,angle=0]{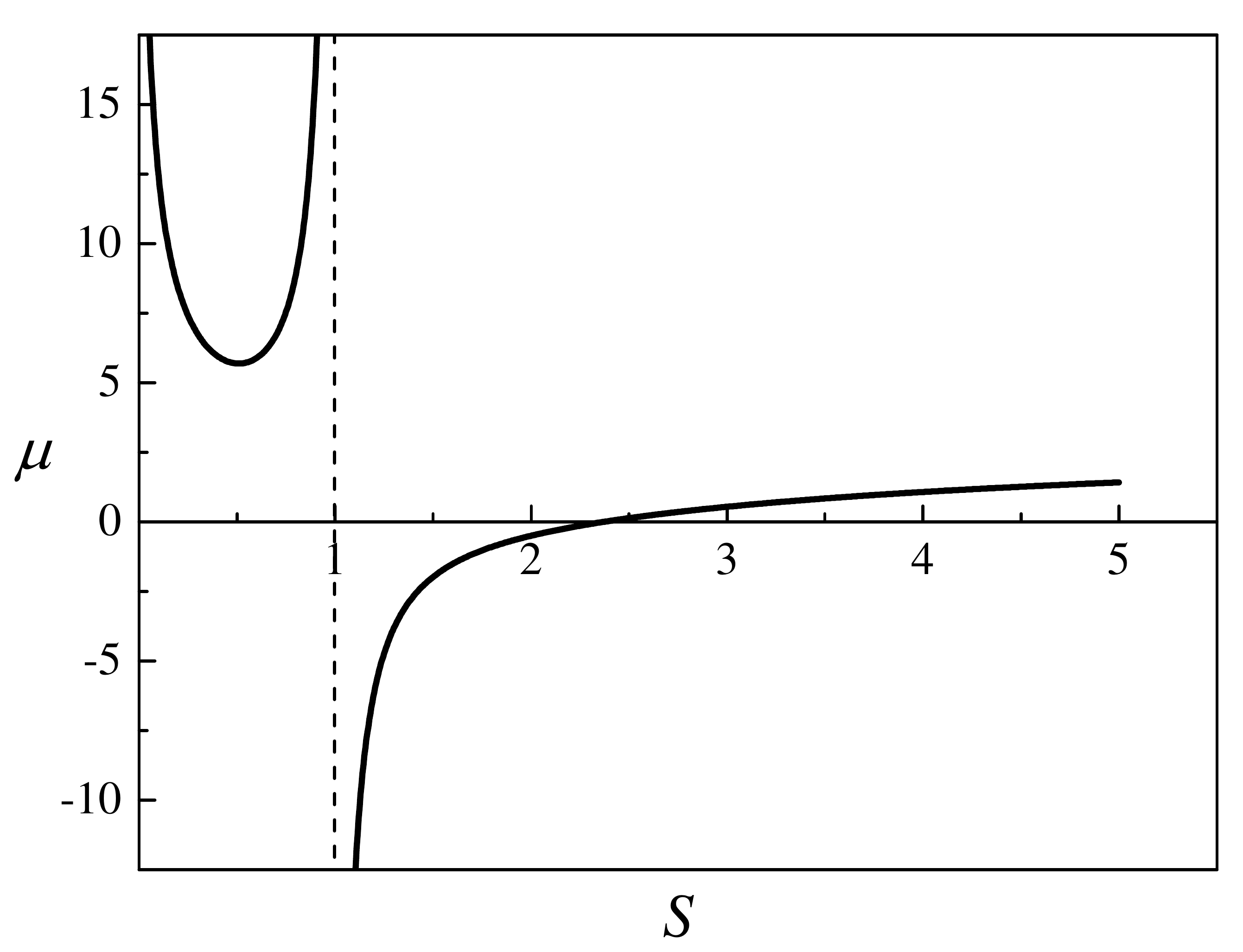}
\end{center}
\caption{The JT coefficient $\mu$ as a function of the KN--AdS black hole entropy $S$, with $p=1$, $J=1$, and $Q=1$. $\mu$ diverges at $S=0.9973$, where the Hawking temperature of the KN--AdS black hole is zero. Then, it monotonically increases and reaches 0 at $S=2.371$. Hence, there are only minimum inversion temperatures for the KN--AdS black holes, and the KN--AdS black holes always cool at large entropies.} \label{muS}
\end{figure}

In Fig. \ref{muS}, we show the JT coefficient of the KN--AdS black hole as a function of entropy $S$. There is a divergent point of $\mu$, where the Hawking temperature of the KN--AdS black hole is zero. Below this entropy, the JT coefficient should be regarded unphysical, as the Hawking temperature is negative. Above this entropy, $\mu$ monotonically increases, but reaches 0 only once, meaning that the heating process happens at smaller entropies and the cooling process happens at larger entropies. As a result, there are only minimum inversion temperatures for the KN--AdS black holes, but no maximum ones, so the KN--AdS black holes always cool at large entropies. This issue will also be discussed in Sects. \ref{sec:inv} and \ref{sec:invc}.

\subsection{Inversion temperature} \label{sec:inv}

Now, we calculate the inversion temperatures of the KN--AdS black holes. Comparing Eq. (\ref{first}) with the differential form of the Smarr relation in Eq. (\ref{smarr}) and taking into account $\rd M=\rd J=\rd Q=0$ in the throttling process, we have
\begin{align}
S\,\rd T-p\,\rd V-2V\,\rd p+J\,\rd \Omega+\f Q2\,\rd \Phi=0. \n
\end{align}
Therefore, the JT coefficient can be reexpressed as
\begin{align}
\mu=\lt(\f{\p T}{\p p}\rt)_{M,J,Q}=\f 1S\lt[p\lt(\f{\p V}{\p p}\rt)_{M,J,Q}-J\lt(\f{\p \Omega}{\p p}\rt)_{M,J,Q}
-\f Q2\lt(\f{\p \Phi}{\p p}\rt)_{M,J,Q}+2V\rt]. \n
\end{align}
Thus, when $\mu=0$, we find the equation of inversion curve,
\begin{align}
p=\lt(\f{\p p}{\p V}\rt)_{M,J,Q}\lt[J\lt(\f{\p \Omega}{\p p}\rt)_{M,J,Q}+\f Q2\lt(\f{\p \Phi}{\p p}\rt)_{M,J,Q}-2V\rt]. \label{pp}
\end{align}
On the other hand, solving Eq. (\ref{M}), we also obtain the pressure in terms of $M$, $J$, $Q$, and $S$,
\begin{align}
p=\f{3}{8S}\lt(2\sqrt{\f{\pi M^2}{S}+\f{\pi^3J^2Q^2}{S^3}+\f{\pi^4J^4}{S^4}}-1-\f{\pi Q^2}{S}-\f{2\pi^2J^2}{S^2}\rt). \label{ppp}
\end{align}
Using Eqs. (\ref{V})--(\ref{Phi}) and (\ref{pp}), and combining the result with Eq. (\ref{ppp}), we may expect to obtain entropy $S$ as a function of $p$, $J$, and $Q$. Substituting $S(p,J,Q)$ into Eq. (\ref{T}), we may finally arrive at the inversion temperature and inversion curve in principle. This procedure is physically straightforward but mathematically hopeless in the present circumstance. The partial derivatives in Eq. (\ref{pp}) are rather tedious for the KN--AdS black holes, because $S$ in Eqs. (\ref{V})--(\ref{Phi}) are also functions of $p$, and $({\p S}/{\p p})_{M,J,Q}$ must also be taken into account. We should mention that even if we start from Eq. (\ref{Ti1}), the situation will not be improved either, because we still need to face $({\p S}/{\p p})_{M,J,Q}$. This complexity terminates all the corresponding calculations, and we believe that it is the essential mathematical obstacle that limits people only to the AdS-like black holes with simple spherical horizon topology and prevents people from discussing the KN--AdS black holes.

Below, we apply a mathematical trick to overcome the difficulty and to greatly simplify the relevant calculations. First, we can reexpress the inversion temperature in Eq. (\ref{Ti1}) as
\begin{align}
T=V\lt(\f{\p T}{\p V}\rt)_{p,J,Q}=V\f{(\p T/\p S)_{p,J,Q}}{(\p V/\p S)_{p,J,Q}}. \label{Ti2}
\end{align}
Substituting Eqs. (\ref{T}) and (\ref{V}) into (\ref{Ti2}), we directly obtain the inversion temperature,
\begin{footnotesize}
\begin{align}
T_{\rm i}&=\f{6\pi ^2 J^2+S[3\pi Q^2+S(3+8pS)]}{12\sqrt{\pi S^3\{12\pi^2J^2(3+8pS)+[3\pi Q^2+S(3+8pS)]^2\}}}\times \n\\
&\quad \f{144\pi^4J^4(9+32pS)+[3\pi Q^2+S(3+8pS)]^3[3\pi Q^2-S(1-8pS)]+24\pi^2J^2[3\pi^2Q^4(9+16pS)+36\pi Q^2S(1+4pS)+S^2(3+8pS)^2(3+16pS)]}
{72\pi^4J^4+6\pi^2J^2[3\pi^2Q^4+2\pi Q^2 S(9+8pS)+3S^2(3+8pS)^2]+S[3\pi Q^2+S(3+8pS)]^3}. \label{Ti3}
\end{align}
\end{footnotesize}However, at the present step, $T_{\rm i}$ is still expressed in terms of $p$, $J$, $Q$, and $S$. Therefore, in order to express the inversion temperature as $T_{\rm i}=T_{\rm i}(p,J,Q)$, we still need to figure out the relation between $S$ and $p$, $J$, and $Q$.

This relation can be obtained by another mathematical trick. We can rewrite Eq. (\ref{Ti1}) as
\begin{align}
0=T-V\lt(\f{\p T}{\p V}\rt)_{p,J,Q}=T^2\lt(\f{\p(V/T)}{\p V}\rt)_{p,J,Q}=T^2\f{(\p(V/T)/\p S)_{p,J,Q}}{(\p V/\p S)_{p,J,Q}}. \n
\end{align}
As a physical black hole always satisfies $T\neq 0$ and $(\p V/\p S)_{p,J,Q}\neq \infty$, what we need is only to set the numerator as
\begin{align}
\lt(\f{\p(V/T)}{\p S}\rt)_{p,J,Q}=0.\n
\end{align}
This trick significantly facilitates the calculations, because the masses $M$ in Eqs. (\ref{T}) and (\ref{V}) cancel out each other in the expression $V/T$ before differentiation. Hence, we easily obtain the relation between $S$ and $p$, $J$, and $Q$,
\begin{align}
&512 p^3 S^7+ 512 p^2 S^6+ 24(7+8\pi pQ^2)p S^5 + 6(3-128\pi^2p^2J^2+16 \pi pQ^2 )S^4 +3\pi(3Q^2-224\pi p J^2-24\pi pQ^4)S^3 \n\\
&-36\pi^2(4 J^2+ Q^4)S^2-27\pi^3Q^2(4 J^2+ Q^4)S- 36 \pi^4J^2(4 J^2+ Q^4) =0.\label{constraint}
\end{align}
It is direct to check that Eq. (\ref{constraint}) reduces to the special results for the RN--AdS black holes when $J=0$ \cite{okcu1} and for the Kerr--AdS black holes when $Q=0$ \cite{okcu2}. Unfortunately, Eq. (\ref{constraint}) is an algebraic equation about $S$ of higher degree, so it is analytically unsolvable. Therefore, we cannot expect to solve $S(p,J,Q)$ directly and substitute it into Eq. (\ref{T}) to obtain the inversion temperature or inversion curve.

We first focus on the minimum inversion temperature $T_{\rm i}^{\rm min}(J,Q)$ (i.e., inversion temperature at vanishing pressure). When $p=0$, Eq. (\ref{Ti3}) reduces to
\begin{align}
T_{\rm i}^{\rm min}=T_{\rm i}^{\rm min}(S,J,Q)=\f{[\pi^2(4J^2+Q^4)(12\pi^2J^2+3\pi^2Q^4+8\pi Q^2 S+6S^2)-S^4](2\pi^2 J^2+\pi Q^2 S+S^2)}
{4\sqrt{\pi S^3[4\pi^2J^2+(\pi Q^2+S)^2]}[8\pi^4J^4+3S(\pi Q^2+S)^3+2\pi^2J^2(\pi Q^2+3S)^2]},\label{Timin}
\end{align}
and Eq. (\ref{constraint}) reduces to
\begin{align}
2S^4 +\pi Q^2S^3-4\pi^2(4 J^2+ Q^4)S^2-3\pi^3Q^2(4 J^2+ Q^4)S- 4\pi^4J^2(4 J^2+ Q^4)=0.\label{22}
\end{align}
Although Eq. (\ref{22}) is still complicated, it is already a quartic equation and is analytically solvable. Among its four solutions, only one is positive and physically meaningful,
\begin{align}
S=S(J,Q)=\f{\pi}{24}\lt[\sqrt{24A+B+\f{C}{A}}+\sqrt{\f{54Q^2(256J^2+63Q^4)}{\sqrt{24A+B+C/A}}-24A+2B-\f{C}{A}}-3Q^2\rt], \label{S1}
\end{align}
where
\begin{align}
A^3&=-22528J^6-9480J^4Q^4-462J^2Q^8+125Q^{12}-6J(4J^2+Q^4)\sqrt{3(294912J^6+97792J^4Q^4-2957J^2Q^8-2250Q^{12})}, \n\\
B&=3[3Q^4+64(4J^2+Q^4)],\quad C=-24[4J^2(32J^2-17Q^4)-25Q^8]. \n
\end{align}
Substituting Eq. (\ref{S1}) into (\ref{Timin}), we can eventually obtain the exact form of the minimum inversion temperature $T_{\rm i}^{\rm min}(J,Q)$. However, the complete expression is tedious, and we will not show it here.

Now, we discuss two limiting cases. For the RN--AdS black holes ($J=0$), Eqs. (\ref{S1}) and (\ref{Timin}) reduce to
\begin{align}
S=\f{3}{2}\pi Q^2,\quad T_{\rm i}^{\rm min}=\f{1}{6\sqrt{6}\pi Q}, \label{limiting}
\end{align}
and we recover the results in Ref. \cite{okcu1}. For the Kerr--AdS black holes ($Q=0$), Eqs. (\ref{S1}) and (\ref{Timin}) reduce to
\begin{align}
S=\sqrt{4+2\sqrt{6}}\pi J,\quad T_{\rm i}^{\rm min}=\f{\sqrt{3}}{4\sqrt[4]{916+374\sqrt{6}}\pi\sqrt{J}}, \n
\end{align}
and we recover the results in Ref. \cite{okcu2}.

Since the complete expression of $T_{\rm i}^{\rm min}(J,Q)$ with finite $J$ and $Q$ is tedious, one may wonder if there is at least a series expansion for it at the neighborhood of $J=Q=0$. However, the answer is no, because $T_{\rm i}^{\rm min}(J,Q)$ is an indeterminate form at $J=Q=0$. For example, if we assume $J=Q^2=\ve\to0$, Eq. (\ref{Timin}) reduces to $T_{\rm i}^{\rm min}={0.01707}/{\sqrt{\ve}}$. Here, we only show the numerical coefficient, as the exact form is very lengthy. However, if we choose another way for $J$ and $Q$ to tend to 0, we may arrive at a different result instead. Therefore, the series expansion of $T_{\rm i}^{\rm min}(J,Q)$ at the neighborhood of $J=Q=0$ is impossible in principle.

For these reasons, it will be highly practical to have an accurate fitting formula for $T_{\rm i}^{\rm min}(J,Q)$, rather than its tedious exact solution. Following Ref. \cite{liu}, we can fit $T_{\rm i}^{\rm min}(J,Q)$ in the parametrization form of
\begin{align}
T_{\rm i}^{\rm min}(J,Q)=\f{\xi(x)}{Q}, \label{fit}
\end{align}
with $x=J/Q^2$. The fitting function $\xi(x)$ is chosen as a rational form,
\begin{align}
\xi(x)=\frac{7.456x^{7/2}-0.4550x^3+3.951x^{5/2}+2.010x^2+1.074x^{3/2} +1.082x+0.1597x^{1/2}+0.3883}
{354.1x^4-23.57x^{7/2}+263.0x^3+78.99x^{5/2}+117.8x^2+54.61 x^{3/2}+47.25x+7.709 x^{1/2}+17.93}.\label{fitt}
\end{align}
The detailed numerical calculations show that the residual sum of squares in this fitting form is only $O(10^{-11})$. Moreover, it reproduces the minimum inversion temperatures of the RN--AdS or Kerr--AdS black holes, when $x\to 0$ or $\infty$.

Below, we compare the minimum inversion temperature $T_{\rm i}^{\rm min}$ with the critical temperature $T_{\rm c}$ of the KN--AdS black holes. If $T_{\rm i}^{\rm min}<T_{\rm c}$, the phase transitions of the KN--AdS black holes in the throttling process are allowed. These comparisons were already performed in the limiting cases: for the RN--AdS black holes, $T_{\rm i}^{\rm min}/T_{\rm c}=1/2$ \cite{okcu1}; for the Kerr--AdS black holes, $T_{\rm i}^{\rm min}/T_{\rm c}=0.5046$ \cite{okcu2}. Both ratios are pure numbers, independent of charges or angular momenta. For the KN--AdS black holes, the critical temperature $T_{\rm c}$ was obtained in Ref. \cite{liu},
\begin{align}
T_{\rm c}=\f{\gamma(x)}{Q}, \quad \gamma(x)=\frac{14.78x^{7/2}-0.9838x^3+7.722x^{5/2}+3.548x^2+2.544x^{3/2}+2.030x+0.3348x^{1/2}+0.7765}
{354.1x^4-23.57x^{7/2}+263.0x^3+78.99x^{5/2}+117.8x^2+54.61x^{3/2}+47.25x+7.709x^{1/2}+17.93}.\label{fittt}
\end{align}
Actually, in Eq. (\ref{fitt}), we have adopted the same denominator for convenience. From Eqs. (\ref{fit})--(\ref{fittt}), we have
\begin{align}
\f{T_{\rm i}^{\rm min}}{T_{\rm c}}=\f{7.456x^{7/2}-0.4550x^3+3.951x^{5/2}+2.010x^2+1.074x^{3/2} +1.082x+0.1597x^{1/2}+0.3883}
{14.78x^{7/2}-0.9838x^3+7.722x^{5/2}+3.548x^2+2.544x^{3/2}+2.030x+0.3348x^{1/2}+0.7765}.\label{bizhi}
\end{align}
Therefore, we can recover the limiting results: for the RN--AdS black holes ($x\to0$), $T_{\rm i}^{\rm min}/T_{\rm c}=0.5001$; for the Kerr--AdS black holes ($x\to\infty$), $T_{\rm i}^{\rm min}/T_{\rm c}=0.5045$. Both numerical ratios are consistent with the exact results in Refs. \cite{okcu1, okcu2}. For the KN--AdS black holes, from Eq. (\ref{bizhi}), $T_{\rm i}^{\rm min}/T_{\rm c}$ lies in the interval $(0.4997,0.5096)$ and depends on the specific values of $J$ and $Q$, but the dependence is not very sensitive. Altogether, the ratios are around $1/2$, and we are convinced that there are possibilities of phase transitions for the KN--AdS black holes in the throttling process.

Here, we should also point out that there are no maximum inversion temperatures for the KN--AdS black holes. This is rather different from the van der Waals-like fluids, and the discrepancy is basically due to the difference in their equations of state. For the van der Waals-like fluids, the second virial coefficient $Tb-a$ depends on temperature. However, from Eq. (\ref{eos}), the second virial coefficient of the KN--AdS black holes is a pure number. It is just this contrast that induces their totally different inversion behaviors, and we will see it more clearly from the inversion curves in Sect. \ref{sec:invc}.

\subsection{Inversion curve} \label{sec:invc}

\begin{figure}[h]
\begin{center}
\subfigure[]{\includegraphics[width=0.4\linewidth,angle=0]{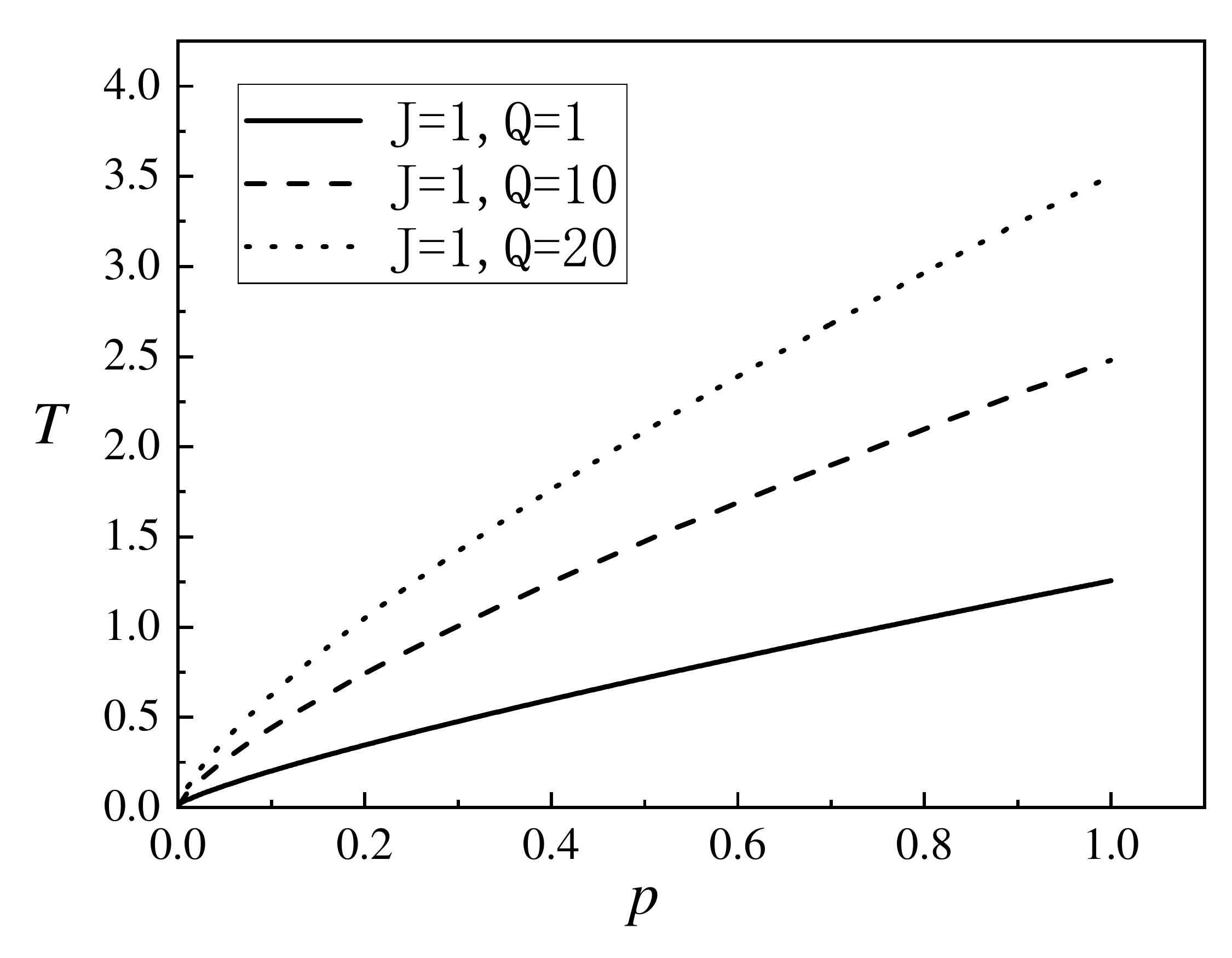} \label{Fig.sub.1}} \quad
\subfigure[]{\includegraphics[width=0.4\linewidth,angle=0]{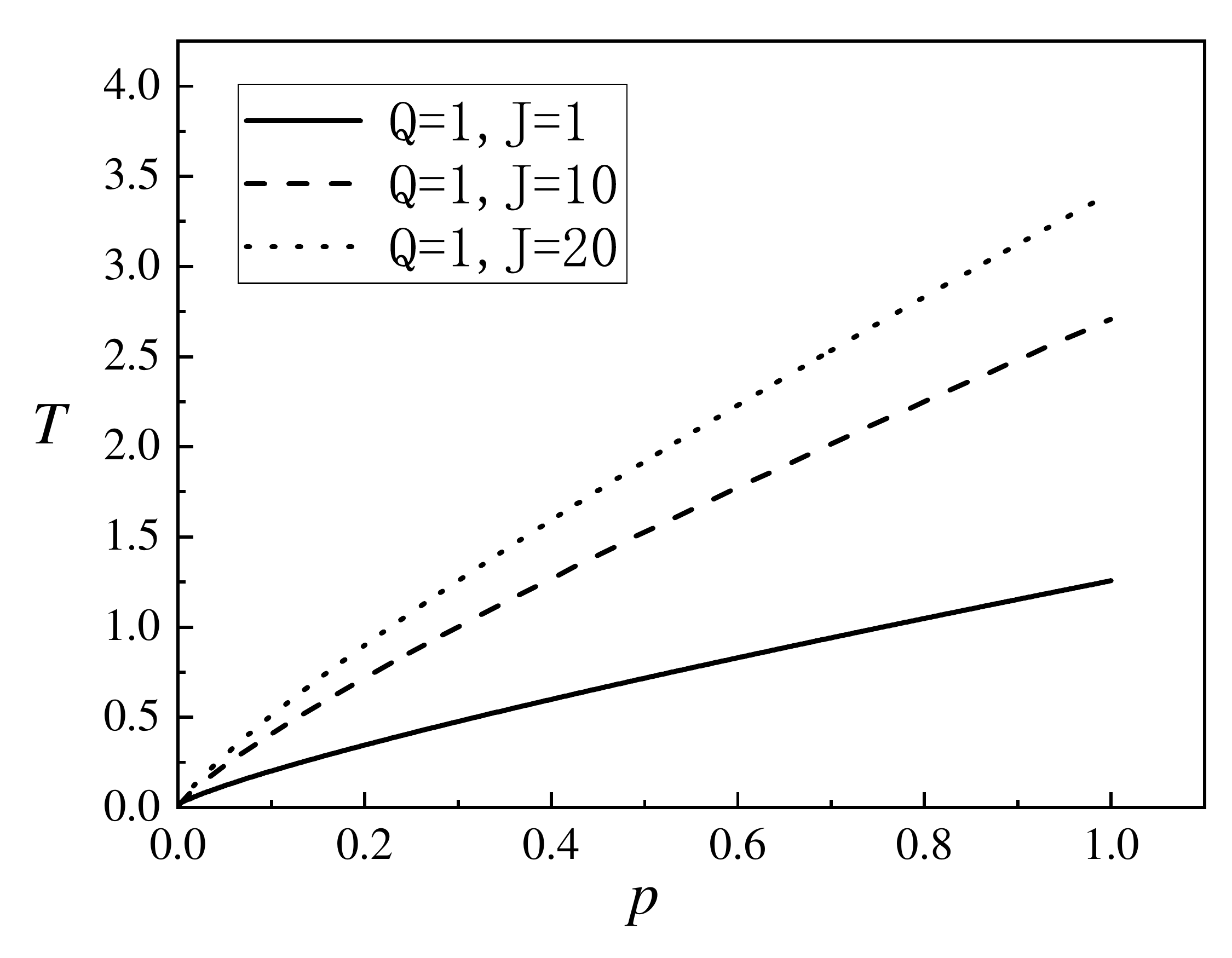} \label{Fig.sub.2}}
\subfigure[]{\includegraphics[width=0.4\linewidth,angle=0]{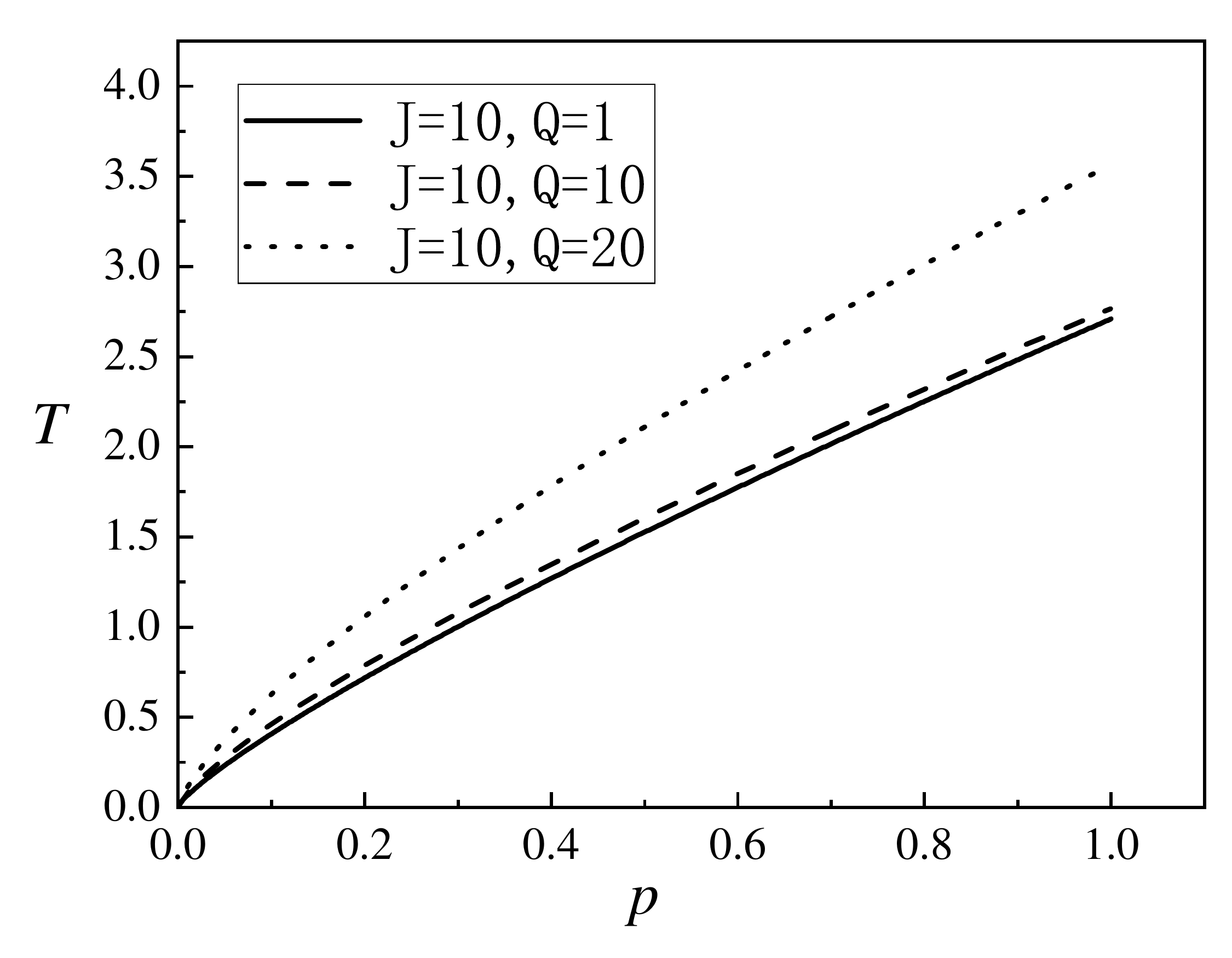} \label{Fig.sub.3}} \quad
\subfigure[]{\includegraphics[width=0.4\linewidth,angle=0]{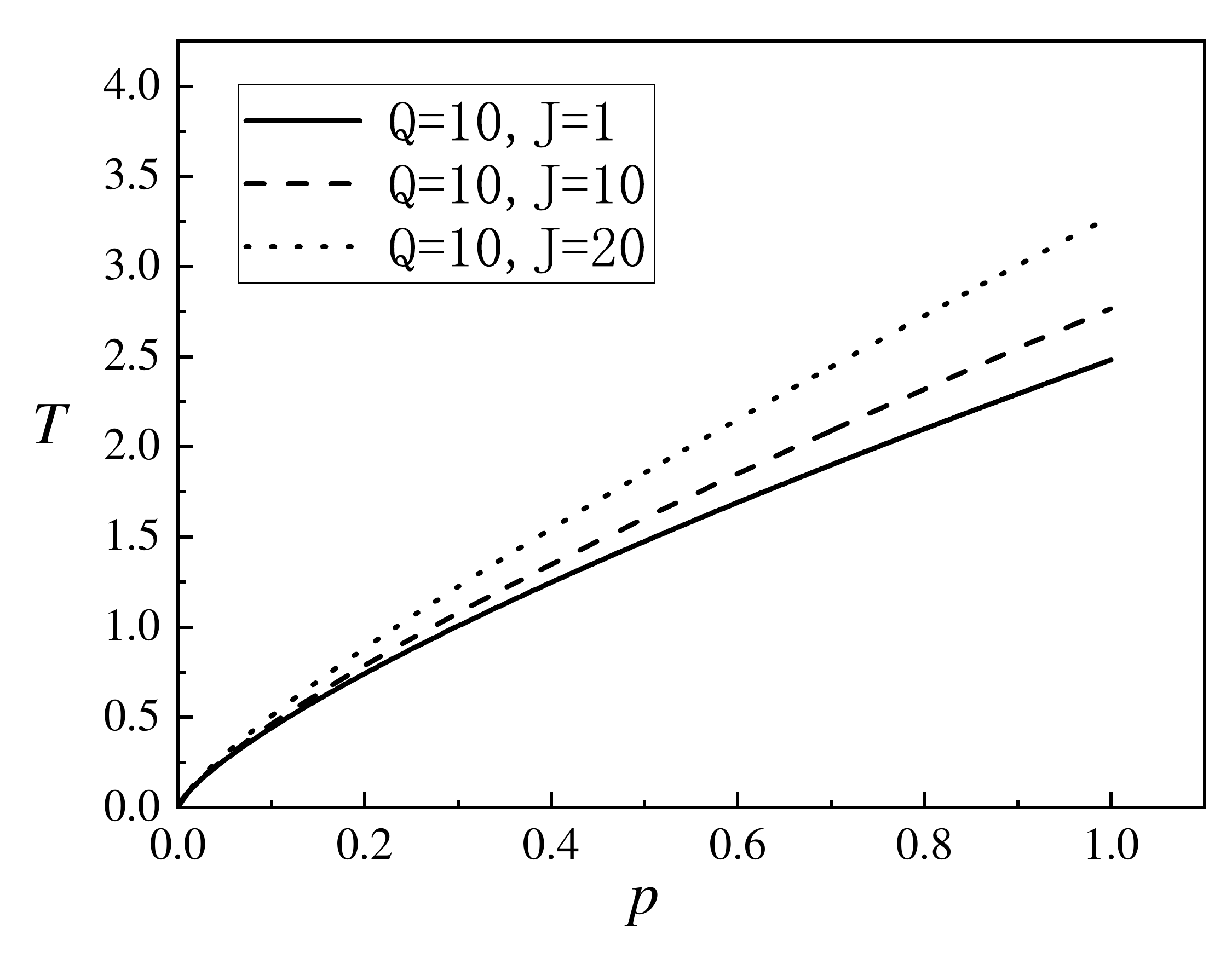} \label{Fig.sub.4}}
\subfigure[]{\includegraphics[width=0.4\linewidth,angle=0]{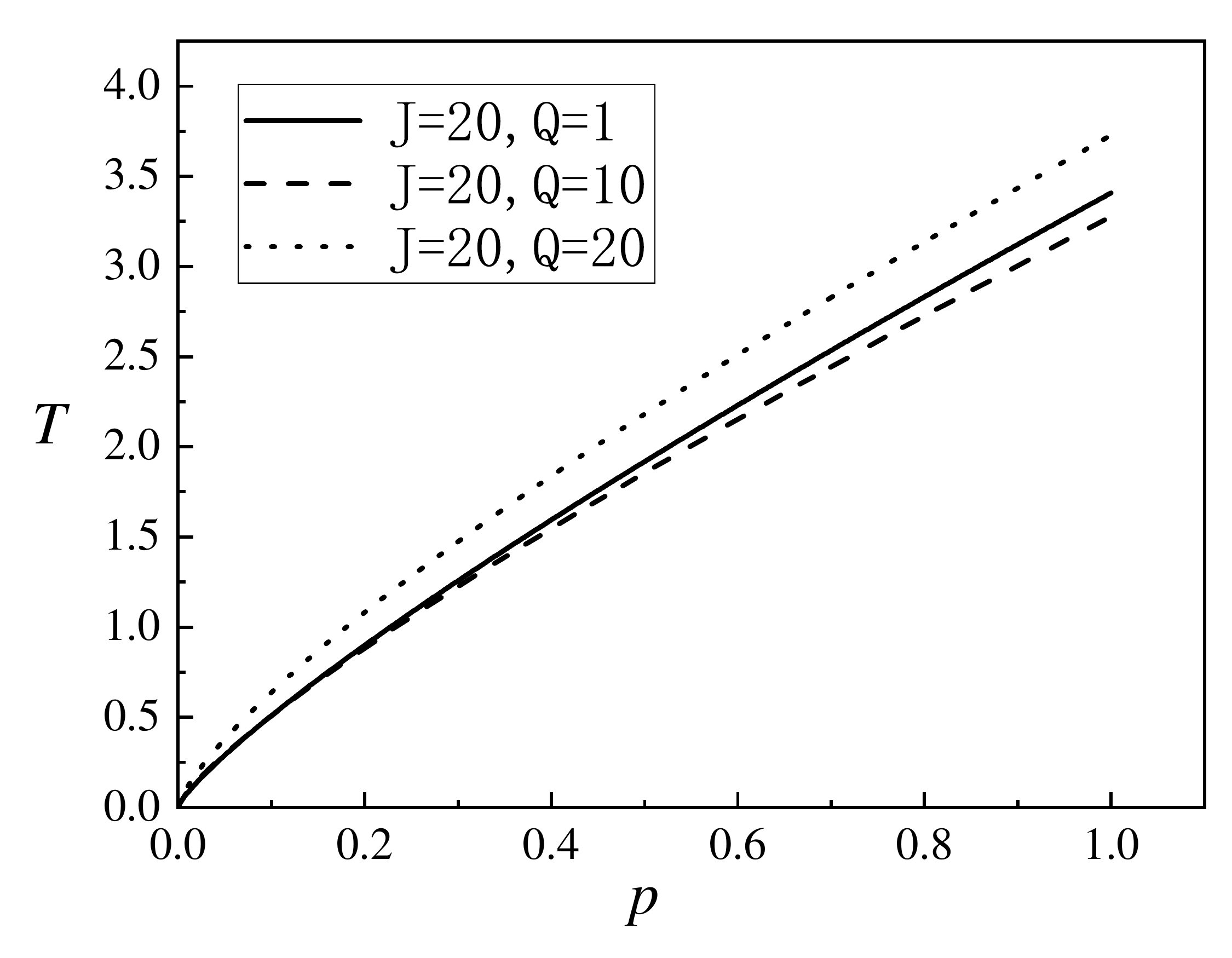} \label{Fig.sub.5}} \quad
\subfigure[]{\includegraphics[width=0.4\linewidth,angle=0]{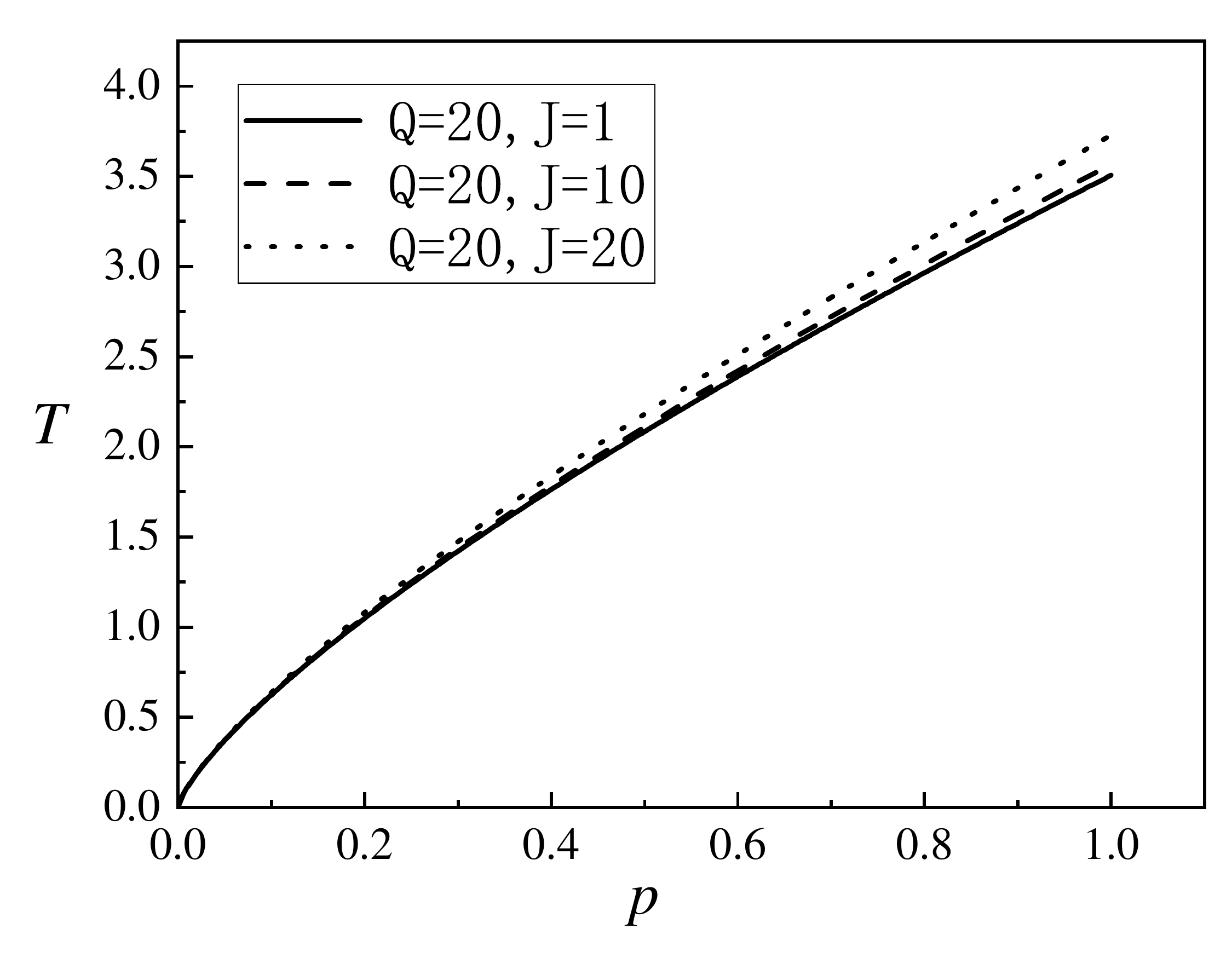} \label{Fig.sub.6}}
\end{center}
\caption{The inversion curves of the KN--AdS black holes with different values of $J$ and $Q$. The figures in the left column show the inversion curves with fixed $J$ and increasing $Q$, and in the right column with fixed $Q$ and increasing $J$. The detailed values of $J$ and $Q$ are listed in each panel. The cooling and heating regions lie above and below the inversion curves respectively. The shapes of the inversion curves are similar, and different inversion curves tend to coincide with each other with $J$ and $Q$ increasing.} \label{Tifig}
\end{figure}

Now, we discuss the inversion curves of the KN--AdS black holes, namely the $T_{\rm i}$--$p$ relations. When $p\neq 0$, the analytical solution of Eq. (\ref{constraint}) is unavailable. Therefore, in Fig. \ref{Tifig}, we display the inversion curves with different values of $J$ and $Q$ by numerical methods.

From Fig. \ref{Tifig}, we observe some characteristic features of the inversion curves. First, the shapes of these curves are similar to those of the RN--AdS \cite{okcu1}, the Kerr--AdS \cite{okcu2}, and other various complicated black holes \cite{D'Almeida, Gh, Mo, Chabab, Mo2018, Lan}. This similarity indicates a fact that it is the cosmological constant that determines the shapes of the inversion curves, not charges or angular momenta. Second, the $T_{\rm i}(p)$ functions are monotonically increasing functions, so there are only minimum inversion temperatures, but no maximum ones. The cooling and heating regions lie above and below the inversion curves, with the JT coefficients positive and negative respectively. Third, with large values of $J$ and $Q$, different inversion curves tend to coincide with each other. For example, in Fig. \ref{Fig.sub.6}, when $Q=20$, the inversion curves with $J=1,10,20$ are very near. This is because when $Q$ is large enough, the influences from different $J$ are negligible. So is the situation when $J$ is large enough.

\begin{figure}[h]
\begin{center}
\includegraphics[width=0.45\linewidth,angle=0]{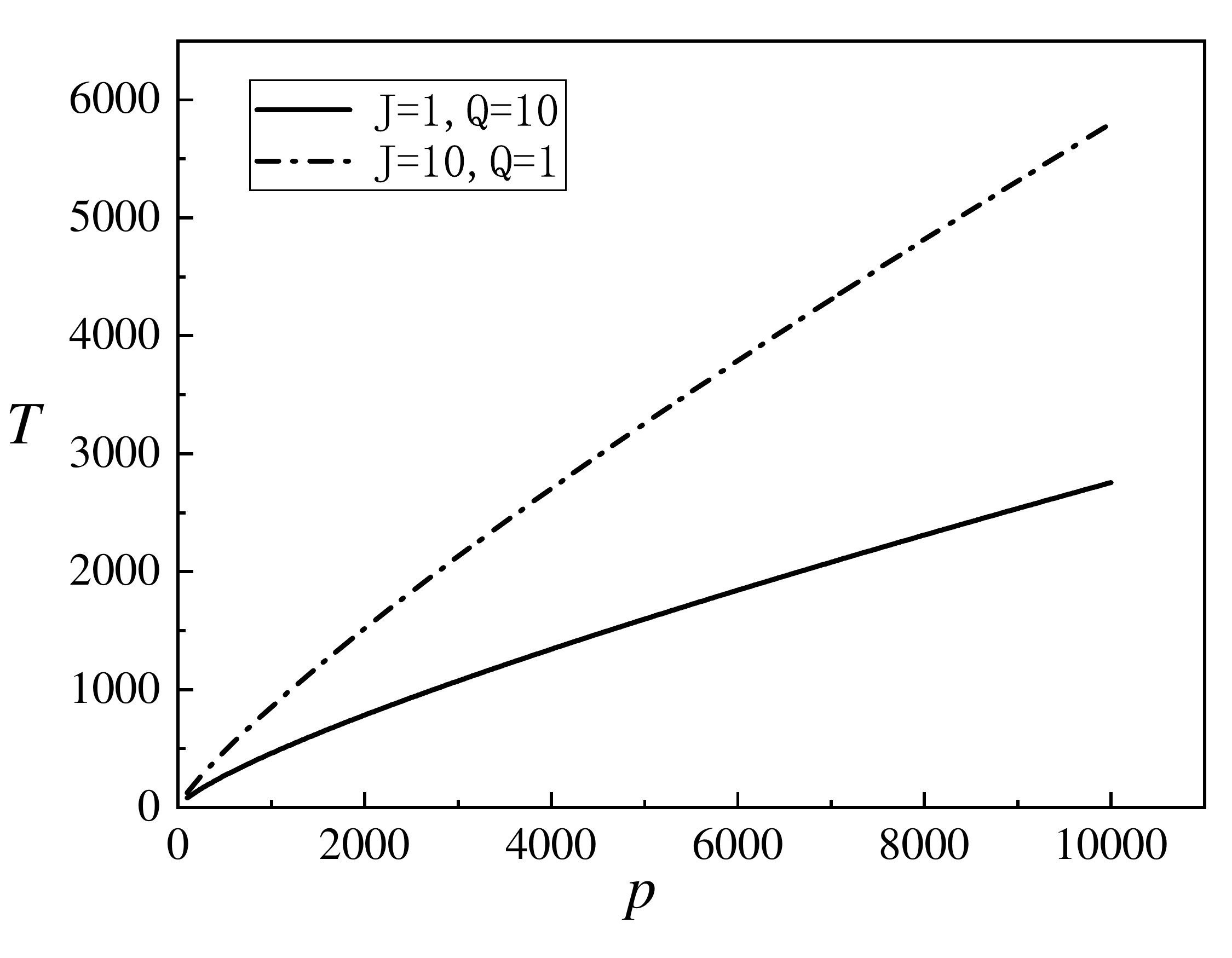}
\end{center}
\caption{The inversion curves of the KN--AdS black holes with different values of $J$ and $Q$ at high pressures ($100<p<10000$). The detailed values of $J$ and $Q$ are listed in the figure. Both inversion curves are concave functions, meaning that there are no maximum inversion temperatures.} \label{pLarge}
\end{figure}

In Fig. \ref{Tifig}, we actually only focus on the inversion curves at low pressures ($0<p<1$). Below, we further discuss their asymptotic behaviors at high pressures ($p\to\infty$). First, for the RN--AdS black holes, Eq. (\ref{constraint}) reduces to $S\propto Q/p^{1/2}$. Substituting it into Eq. (\ref{Ti3}), we have $T_{\rm i}\propto Q^{1/2}p^{3/4}$ at high pressures. These proportions are consistent with the exact results in Ref. \cite{okcu1}. Second, for the Kerr--AdS black holes, $S\propto J^{2/3}/p^{1/3}$ and $T_{\rm i}\propto J^{1/3}p^{5/6}$ at high pressures, and these proportions are also consistent with the numerical results in Ref. \cite{okcu2}. In each case, $T_{\rm i}$ is a power function of $p$, with the power less than 1, meaning that $T_{\rm i}(p)$ is a concave function at large pressures. Similarly, in Fig. \ref{pLarge}, we plot the inversion curves at high pressures for the KN--AdS black holes with different values of $J$ and $Q$ by numerical methods. Again, we observe that the inversion curves are concave functions. This also supplies an evidence that there are no maximum inversion temperatures, so the KN--AdS black holes always cool above the inversion curves.

\subsection{Isenthalpic curve} \label{sec:iso}

\begin{figure}[h]
\begin{center}
\subfigure[]{\includegraphics[width=0.4\linewidth,angle=0]{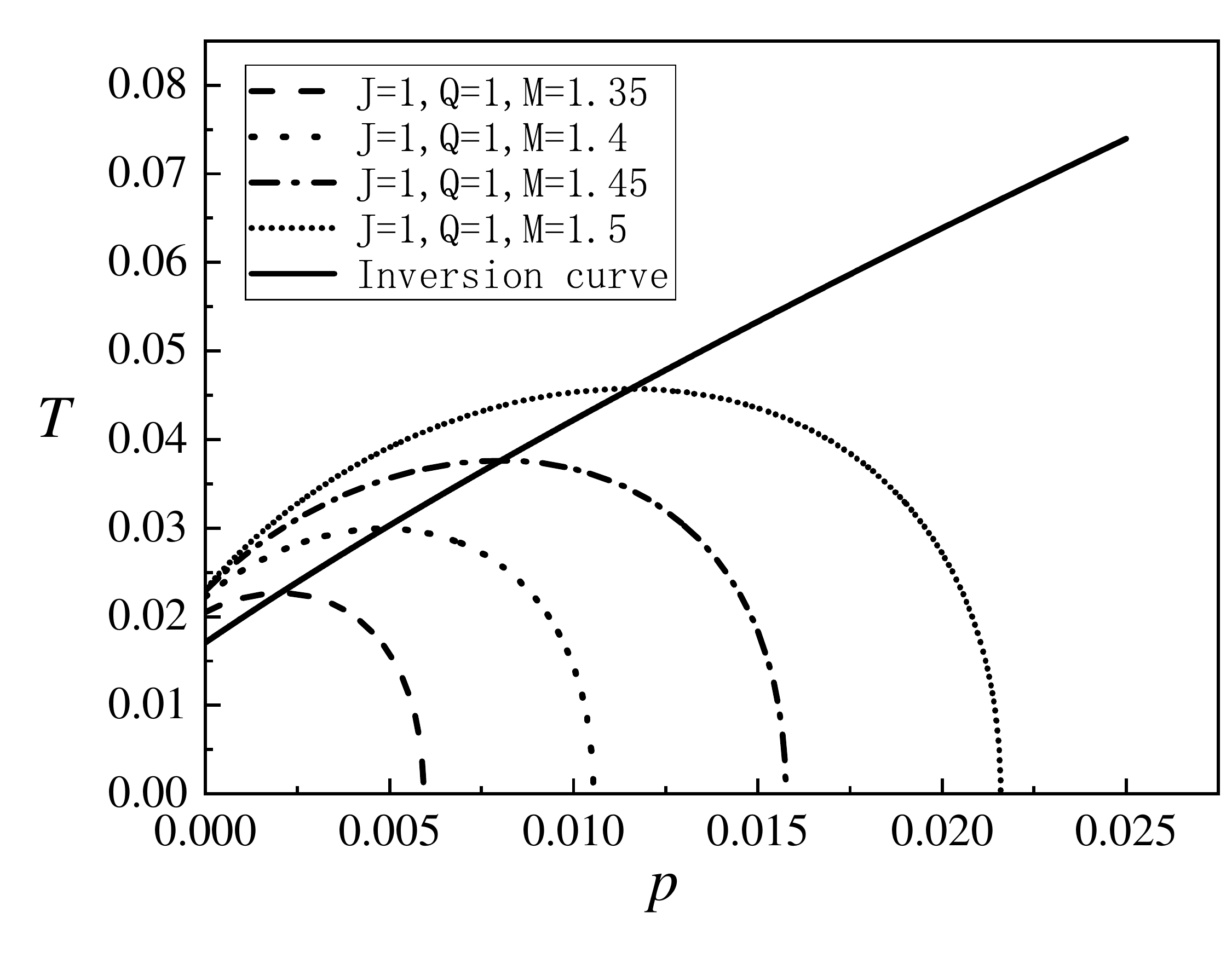} \label{51new}}\quad
\subfigure[]{\includegraphics[width=0.4\linewidth,angle=0]{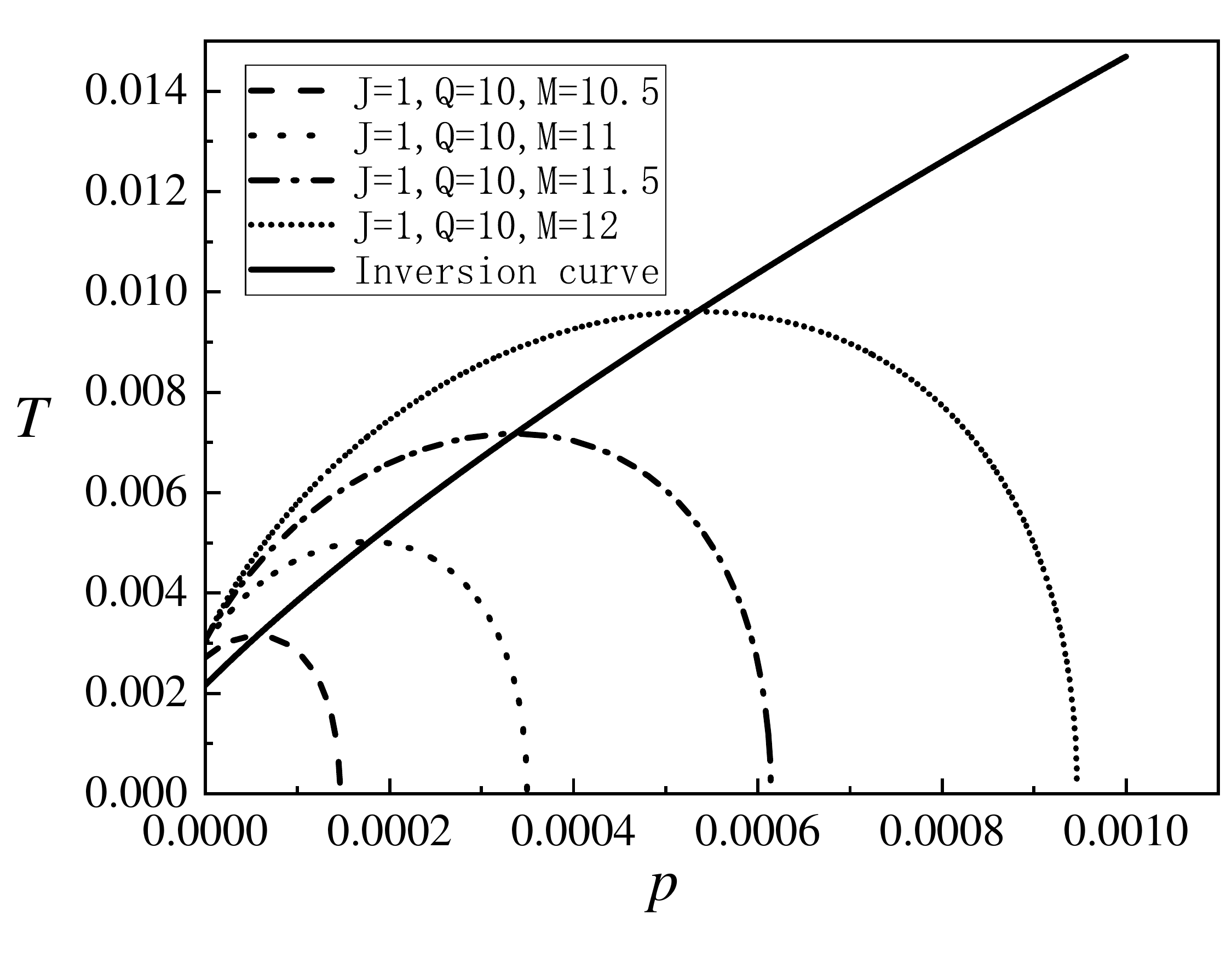} \label{52}}
\subfigure[]{\includegraphics[width=0.4\linewidth,angle=0]{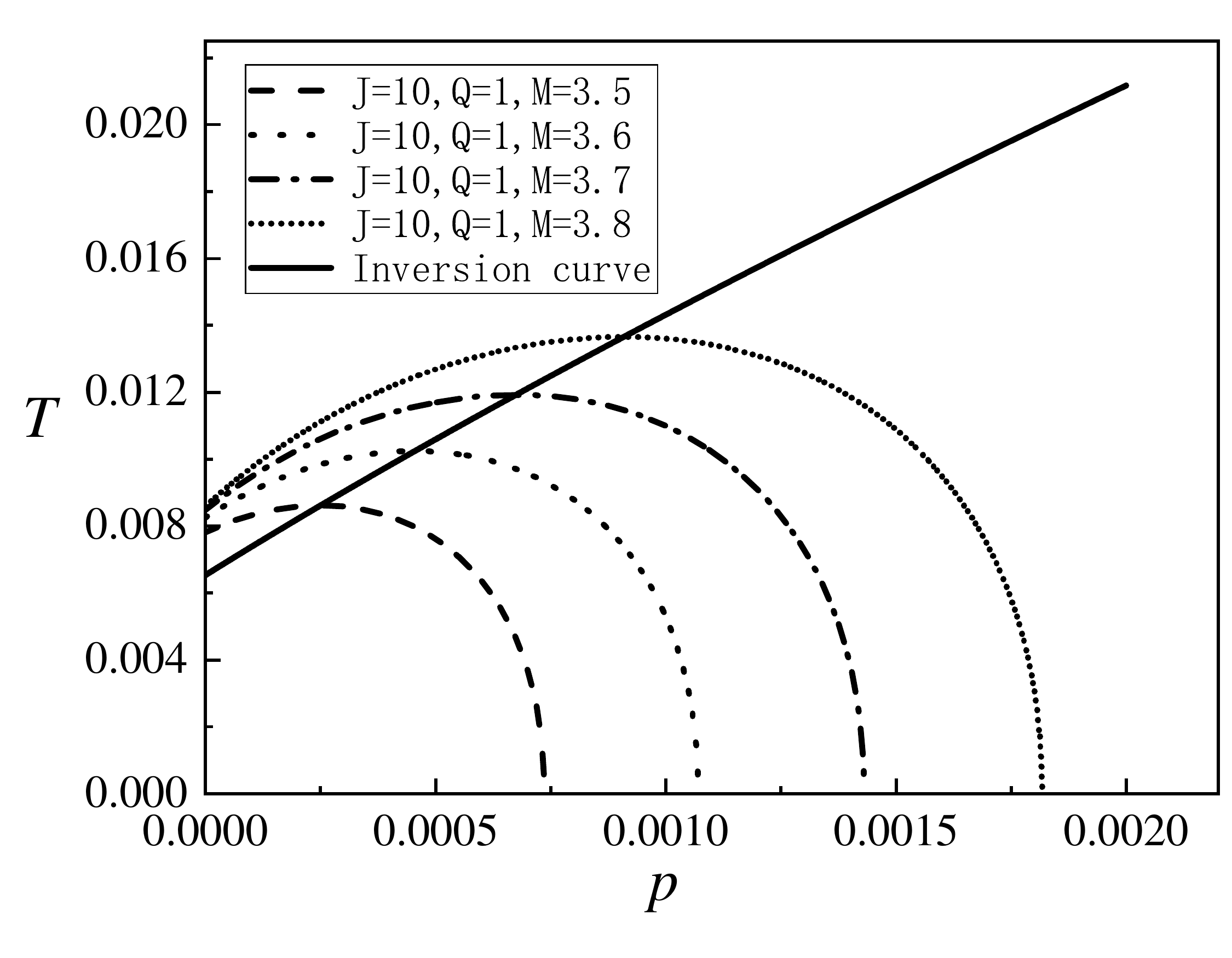} \label{53}} \quad
\subfigure[]{\includegraphics[width=0.4\linewidth,angle=0]{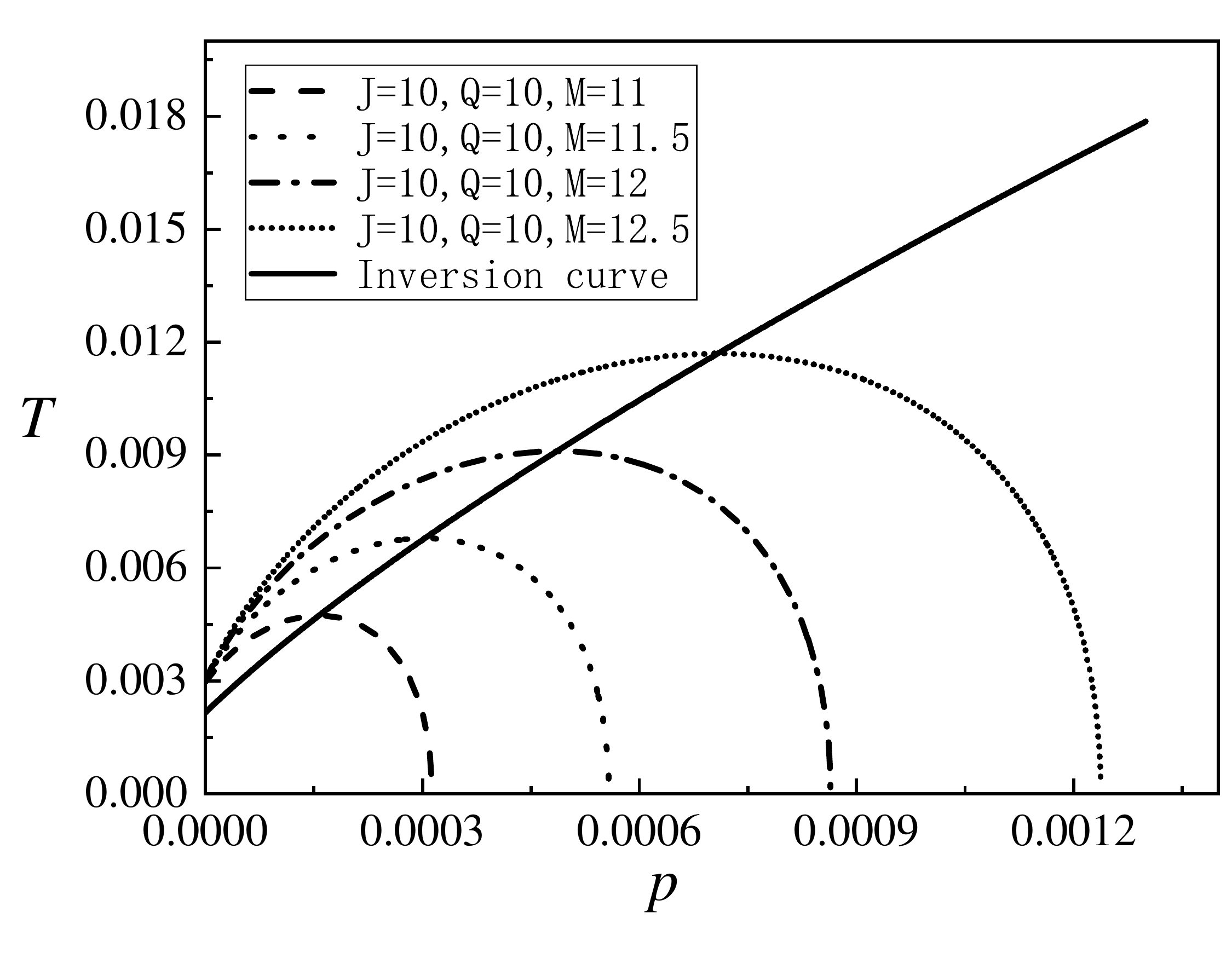} \label{54}}
\end{center}
\caption{The isenthalpic curves and the inversion curves of the KN--AdS black holes with different values of $J$ and $Q$. The detailed values of $J$ and $Q$ are listed in each panel. The shapes of the isenthalpic curves are similar to those of the RN--AdS or Kerr--AdS black holes, with large $Q$ or $J$. In each panel, the masses of the KN--AdS black holes must satisfy the condition $2M^2>\sqrt{4 J^2+Q^4}+Q^2$ to avoid naked singularity.}\label{Q1TiequM}
\end{figure}

As we have explained, the throttling process of a black hole is in fact an isenthalpic (i.e., constant-mass) process in the extended phase space. In Fig. \ref{Q1TiequM}, we show the isenthalpic curves and the corresponding inversion curves of the KN--AdS black holes with different values of $J$ and $Q$ by numerical methods. To avoid naked singularity, the masses of the KN--AdS black holes must be carefully chosen, such that $2M^2>\sqrt{4 J^2+Q^4}+Q^2$. From Fig. \ref{Q1TiequM}, we clearly observe that with large $Q$, the shapes of the isenthalpic curves are similar to those of the RN--AdS black holes in Ref. \cite{okcu1}; with large $J$, the shapes of the isenthalpic curves are similar to those of the Kerr--AdS black holes in Ref. \cite{okcu2}.

Now, we explore two notable issues not much mentioned in the literature: the isenthalpic curves that do not intersect with the inversion curves and the isenthalpic curves that intersect with each other. Two corresponding characteristic masses, $M_\ast$ and $\widetilde{M}$, will also be discussed.

First, we can classify the isenthalpic curves of the KN--AdS black holes into two types by comparing their Hawking temperatures at vanishing pressure, $T_0$, with their minimum inversion temperatures, $T_{\rm i}^{\rm min}$. First, if $T_0>T_{\rm i}^{\rm min}$, the isenthalpic curves rise and then fall as pressure increases (with the slopes being positive and negative respectively), and an inversion curve is just the line that connects the extreme points of these isenthalpic curves. Second, if $T_0<T_{\rm i}^{\rm min}$, the isenthalpic curves monotonically descend, so no inversion curve exists at all. In between, there is a special isenthalpic curve satisfying $T_0=T_{\rm i}^{\rm min}$, with the corresponding black hole mass being $M_\ast$. In other words, its extreme point is located at $(0,T_{\rm i}^{\rm min})$ and thus separates the two types of isenthalpic curves.

\begin{figure}[h]
\begin{center}
\subfigure[]{\includegraphics[width=0.4\linewidth,angle=0]{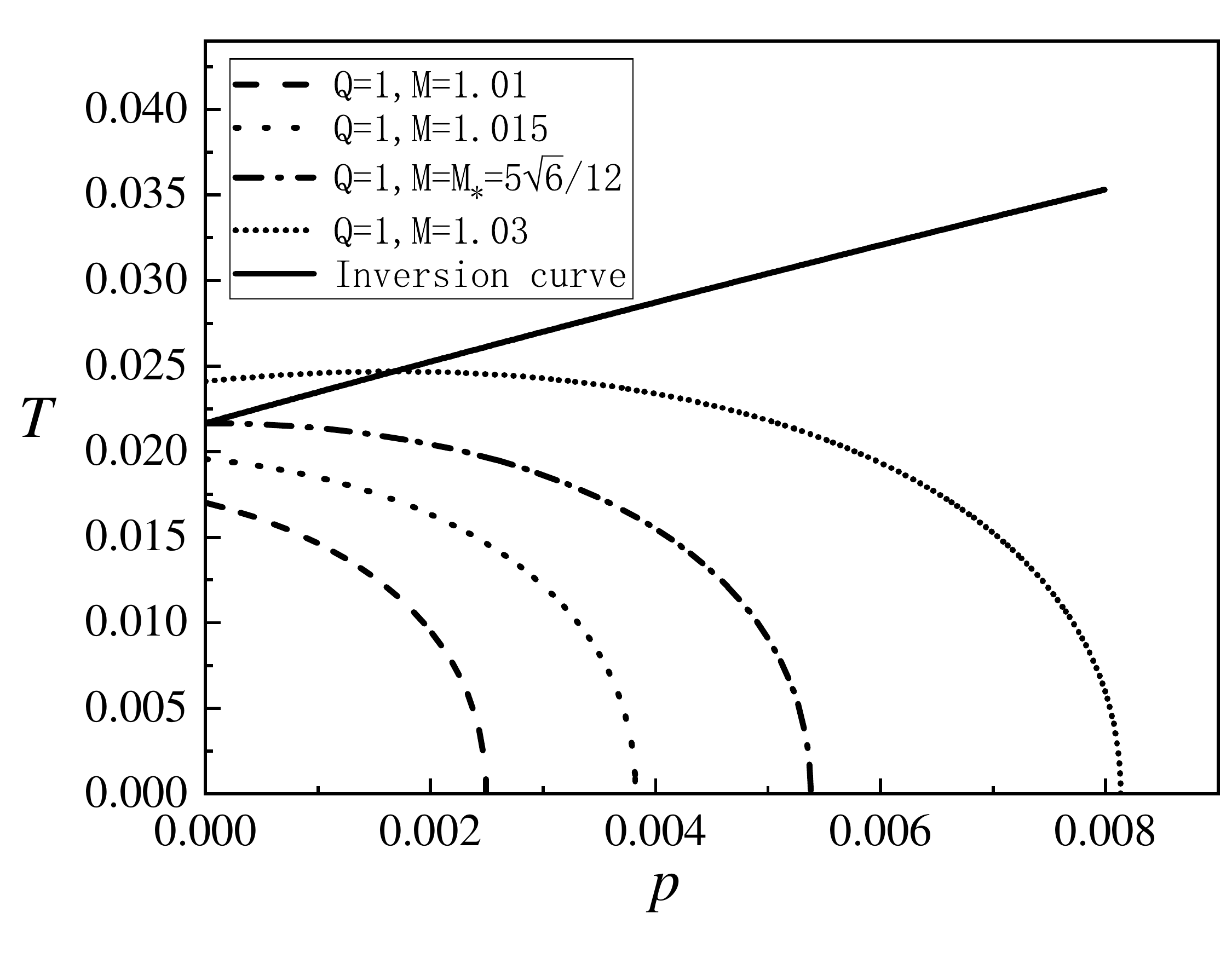} \label{61}} \quad
\subfigure[]{\includegraphics[width=0.4\linewidth,angle=0]{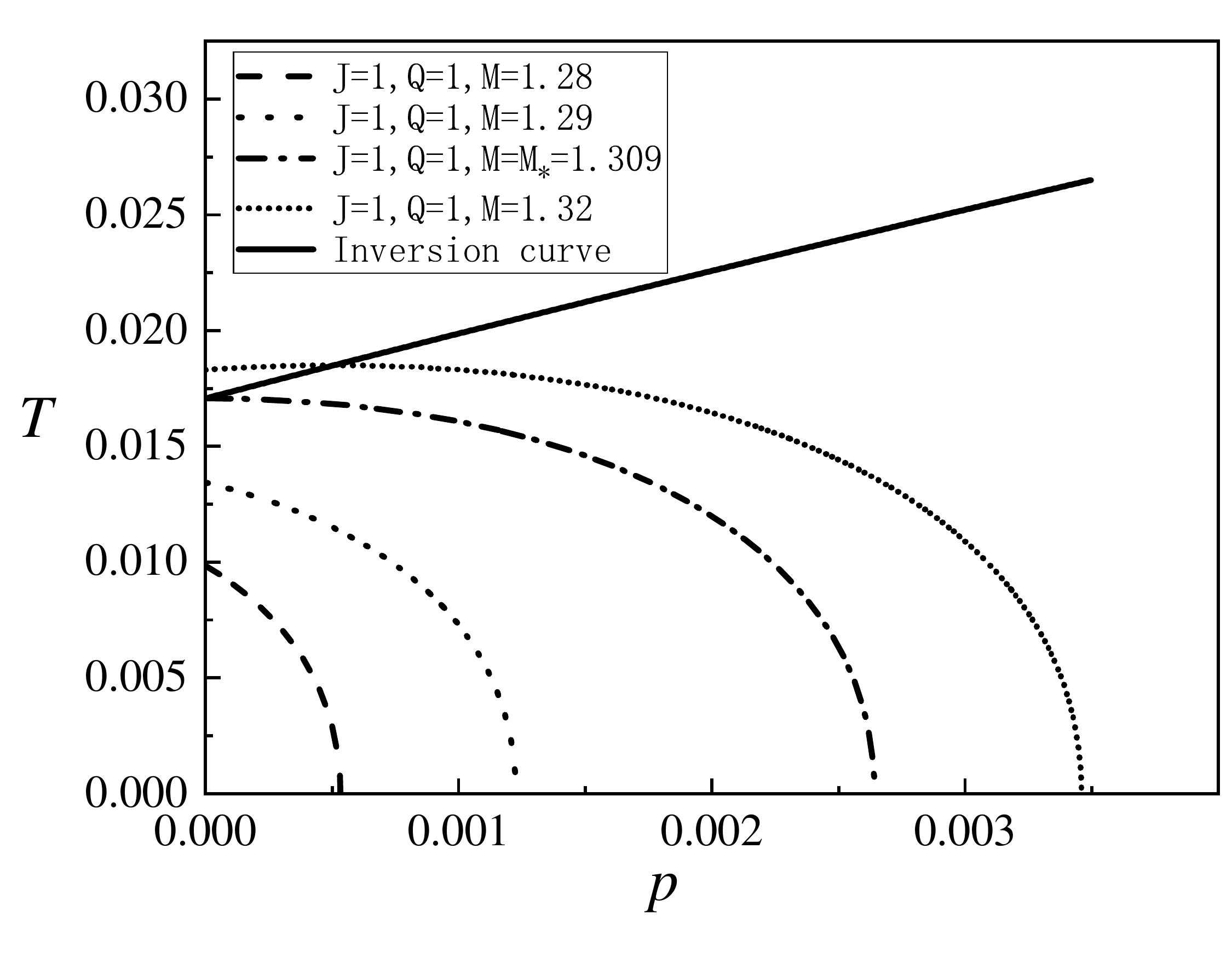} \label{62}}
\end{center}
\caption{The isenthalpic curves and the inversion curves of the RN--AdS (left panel) and KN--AdS (right panel) black holes, with the black hole masses chosen around their characteristic values $M_\ast$. The detailed values of $J$, $Q$, and $M$ are listed in each panel. If $M>M_\ast$, there is an inversion curve; if $M<M_\ast$, there is no inversion curve.}\label{lbf}
\end{figure}

We take the RN--AdS black holes as an example to elucidate this issue explicitly, as this case is analytically solvable. For the RN--AdS black holes, Eqs. (\ref{M}) and (\ref{T}) reduce to
\begin{align}
M=\frac{1}{2} \sqrt{\frac{S}{\pi}} \left(1+\f{\pi Q^2}{S}+\frac{8 p S}{3}\right),\quad T=\f{1}{4\sqrt{\pi S}}\left(1-\f{\pi Q^2}{S}+{8 p S}\right).\n
\end{align}
When $p=0$, from the equations above, we obtain
\begin{align}
T_0=\frac{\sqrt{M^2- Q^2}}{2 \pi(M+\sqrt{M^2- Q^2})^{2}}.\label{Tp0}
\end{align}
Comparing this result with the minimum inversion temperature in Eq. (\ref{limiting}), $T_{\rm i}^{\rm min}=1/(6\sqrt{6}\pi Q)$, we obtain
\begin{align}
M_\ast=\f{5\sqrt{6}Q}{12}. \n
\end{align}
This is the characteristic mass for the isenthalpic curves of the RN--AdS black holes. If $M>M_\ast$, there exists an inversion curve; if $M<M_\ast$, there is no. In Fig. \ref{61}, the characteristic isenthalpic curve and the isenthalpic curves with and without the inversion curve are plotted together. The discussions on the Kerr--AdS and KN--AdS black holes are qualitatively analogous, so we only show the KN--AdS case in Fig. \ref{62} by numerical methods. For the Kerr--AdS black holes, $M_\ast=\sqrt[4]{(2+3\sqrt{6})/8}\sqrt{J}$; for the KN--AdS black holes, $M_\ast=1.309$ (we choose $J=1$ and $Q=1$).

\begin{figure}[h]
\begin{center}
\subfigure[]{\includegraphics[width=0.4\linewidth,angle=0]{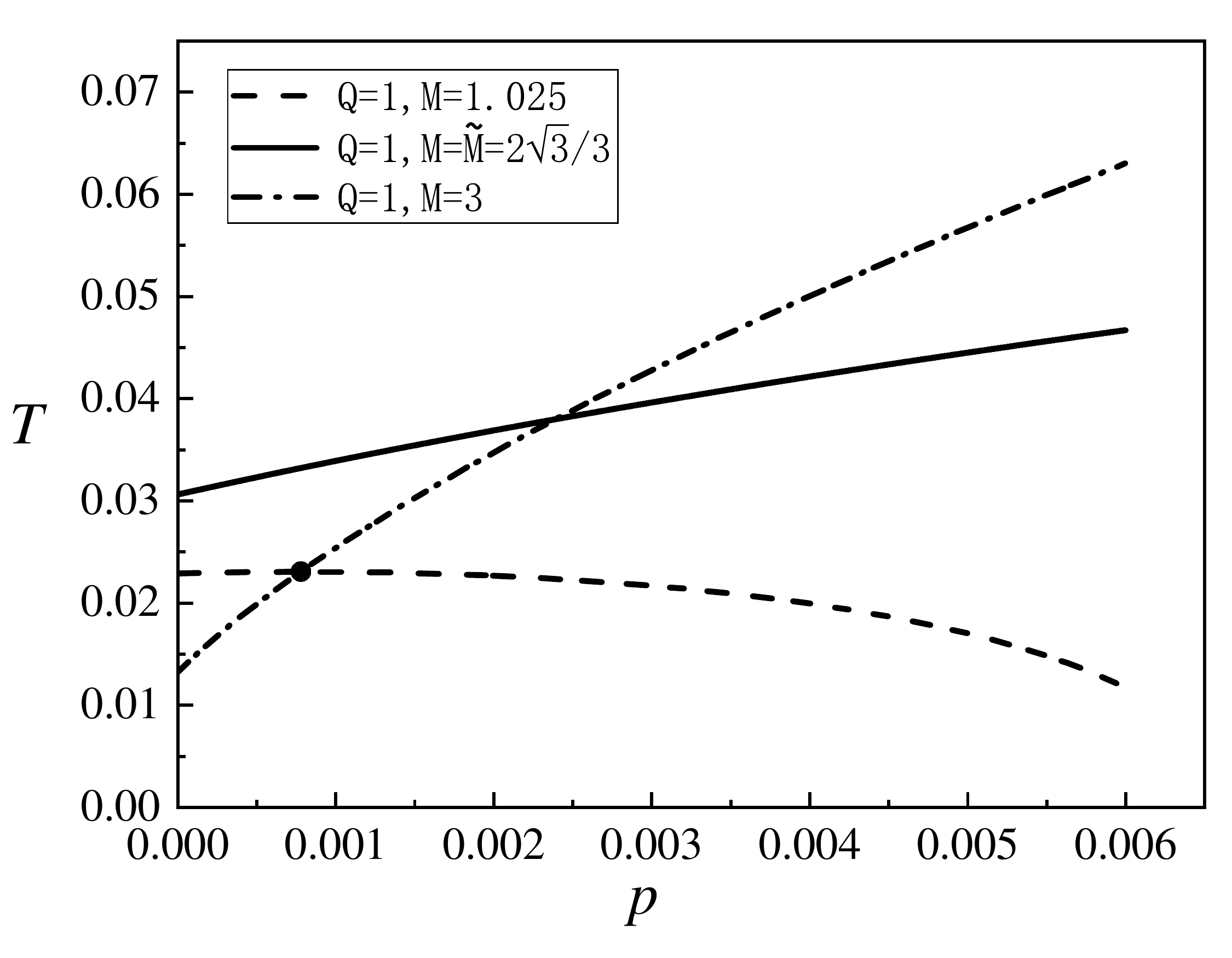} \label{71}} \quad
\subfigure[]{\includegraphics[width=0.4\linewidth,angle=0]{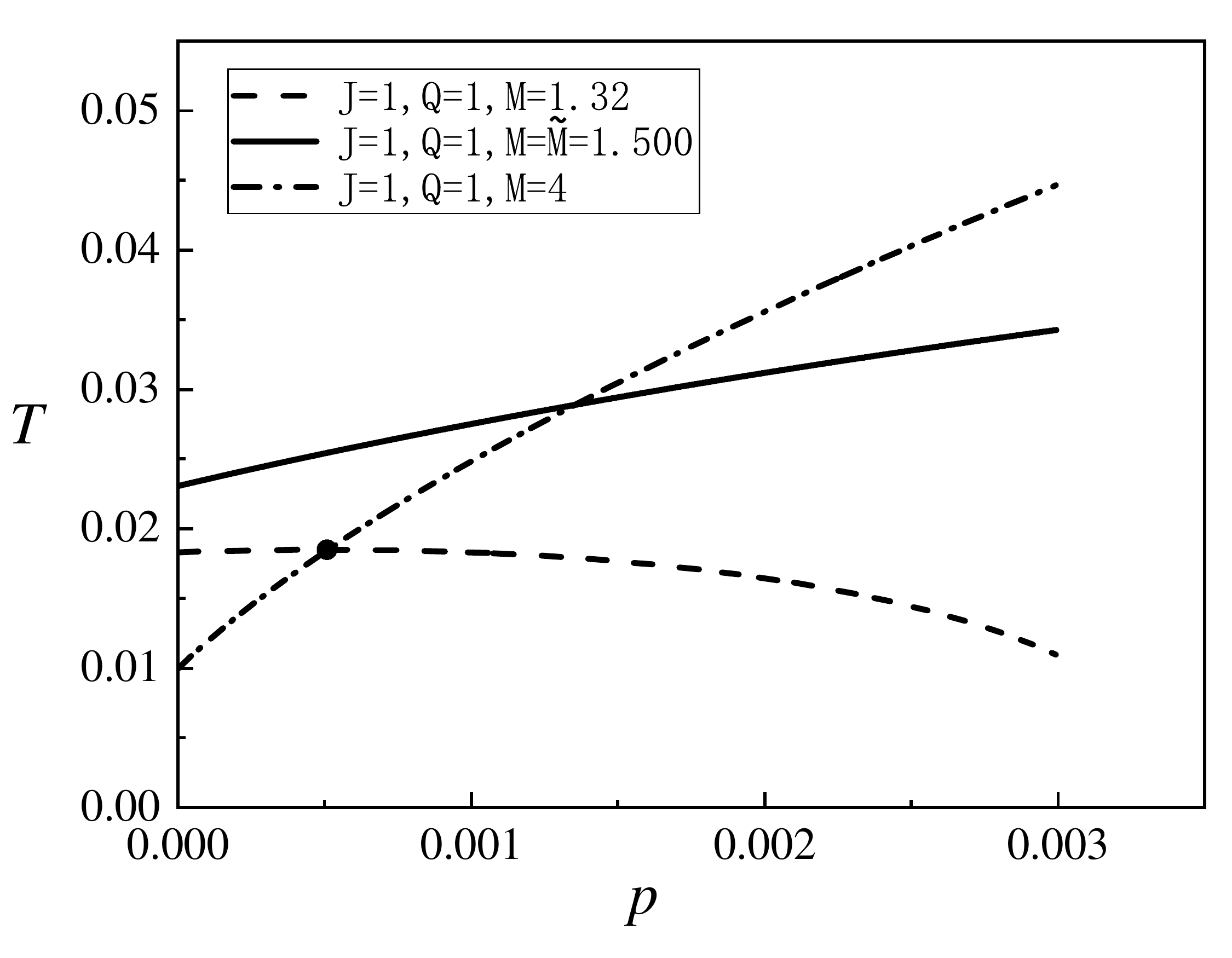} \label{72}}
\end{center}
\caption{The intersections of the isenthalpic curves with different enthalpies (i.e., masses) for the RN--AdS (left panel) and KN--AdS (right panel) black holes. The detailed values of $J$, $Q$, and $M$ are listed in each panel.}\label{Mwan}
\end{figure}

Second, in Fig. \ref{Mwan}, we show the intersections of the isenthalpic curves for the RN--AdS and KN--AdS black holes. One may wonder why the isenthalpic curves with different enthalpies (i.e., masses) may pass the same point in the $T$--$p$ plane. Actually, this is inevitable. For the given values of $T$ and $p$, there may still be different values of $S$ in Eq. (\ref{T}), and the black hole masses are thus multi-valued in Eq. (\ref{M}). For example, the black hole horizon radii $r_+$ are obviously different before and after the small--large black hole phase transitions while keeping $T$ and $p$ constant \cite{KM}. So are the black hole entropies and masses.

\begin{figure}[h]
\begin{center}
\subfigure[]{\includegraphics[width=0.4\linewidth,angle=0]{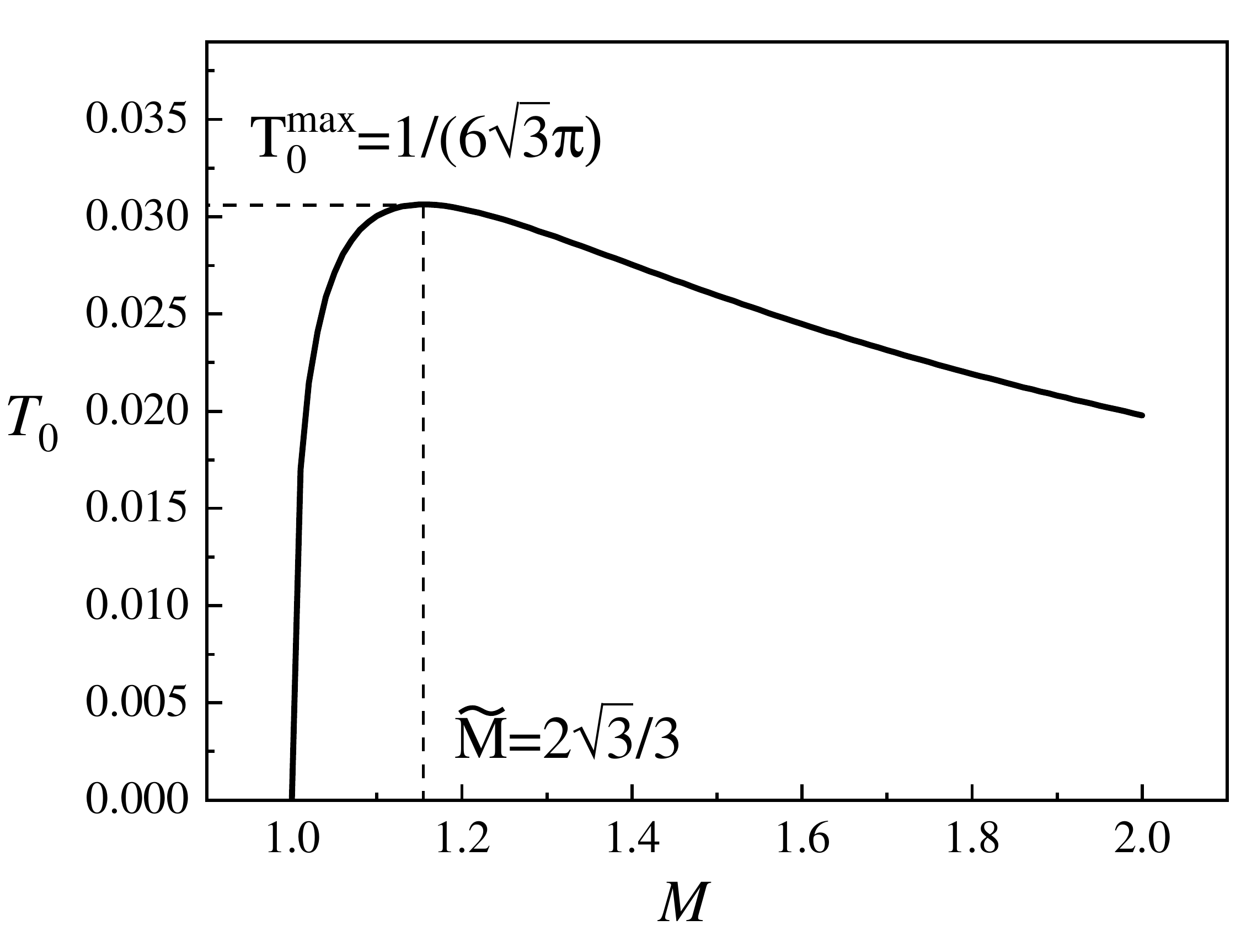} \label{81}} \quad
\subfigure[]{\includegraphics[width=0.4\linewidth,angle=0]{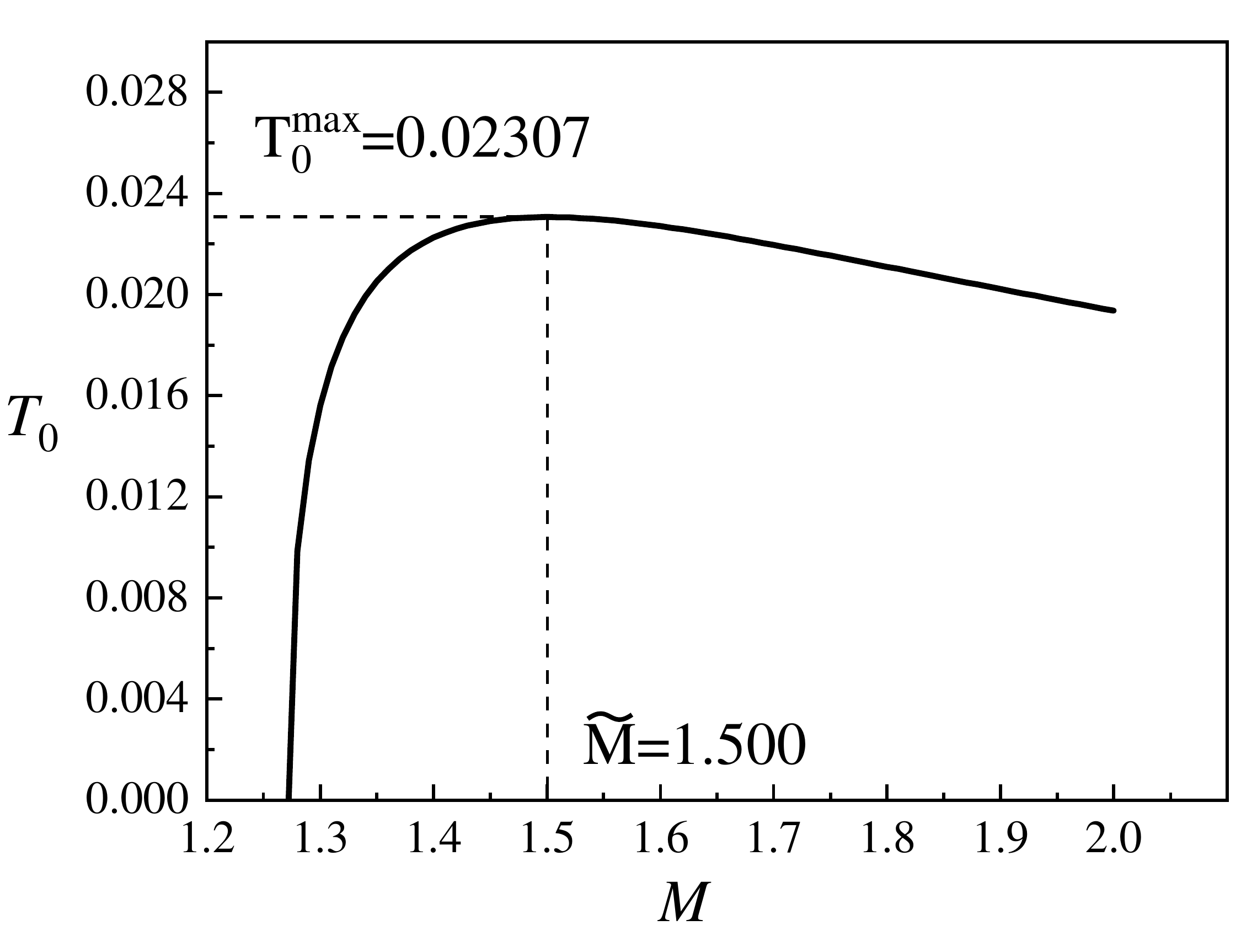} \label{82}}
\end{center}
\caption{The $T$-intercepts of the isenthalpic curves as a function of the black hole masses. For the RN--AdS black holes (left panel), we choose $Q=1$, and the maximum point is located at $({2\sqrt{3}}/{3},{1}/(6\sqrt{3}\pi))$. For the KN--AdS black holes (right panel), we choose $J=1$ and $Q=1$, and the maximum point is located at $(1.500,0.02307)$.} \label{TRN}
\end{figure}

We again take the RN--AdS black holes as an analytical example. From Eq. (\ref{Tp0}), the $T$-intercepts (i.e., $T_0$) of the isenthalpic curves first rise and then fall as $M$ grows. However, the $p$-intercepts keep increasing steadily. Therefore, the leftmost ends of the isenthalpic curves first go up to the highest point $T_0^{\rm max}$ (when $M$ equals a specific value $\widetilde{M}$) and then down, but the rightmost ends keep moving to the right. This unavoidably leads to the intersections of the isenthalpic curves with different masses. Taking the maximum of $T_0$ in Eq. (\ref{Tp0}) yields
\begin{align}
\widetilde{M}=\f{2\sqrt{3}Q}{3}, \quad T_0^{\rm max}=\f{1}{6\sqrt{3}\pi Q}. \n
\end{align}
Hence, an isenthalpic curve with $M>\widetilde{M}$ must intersect with another one in the $T$--$p$ plane, because its $T$-intercept is less than $T_0^{\rm max}$. The cases are again similar for the Kerr--AdS and KN--AdS black holes. For the Kerr--AdS black holes, $\widetilde{M}=\sqrt[4]{(3+2\sqrt{3})/3}\sqrt{J}$ and $T_0^{\rm max}=1/(2\sqrt[4]{540+312\sqrt{3}}\pi\sqrt{J})$; for the KN--AdS black holes, $\widetilde{M}=1.500$ and $T_0^{\rm max}=0.02307$ (we choose $J=1$ and $Q=1$). The $T$-intercepts as a function of $M$ are shown in Fig. \ref{81} for the RN--AdS black holes analytically and in Fig. \ref{82} for the KN--AdS black holes by numerical methods.

\begin{figure}[h]
\begin{center}
\includegraphics[width=0.6\linewidth,angle=0]{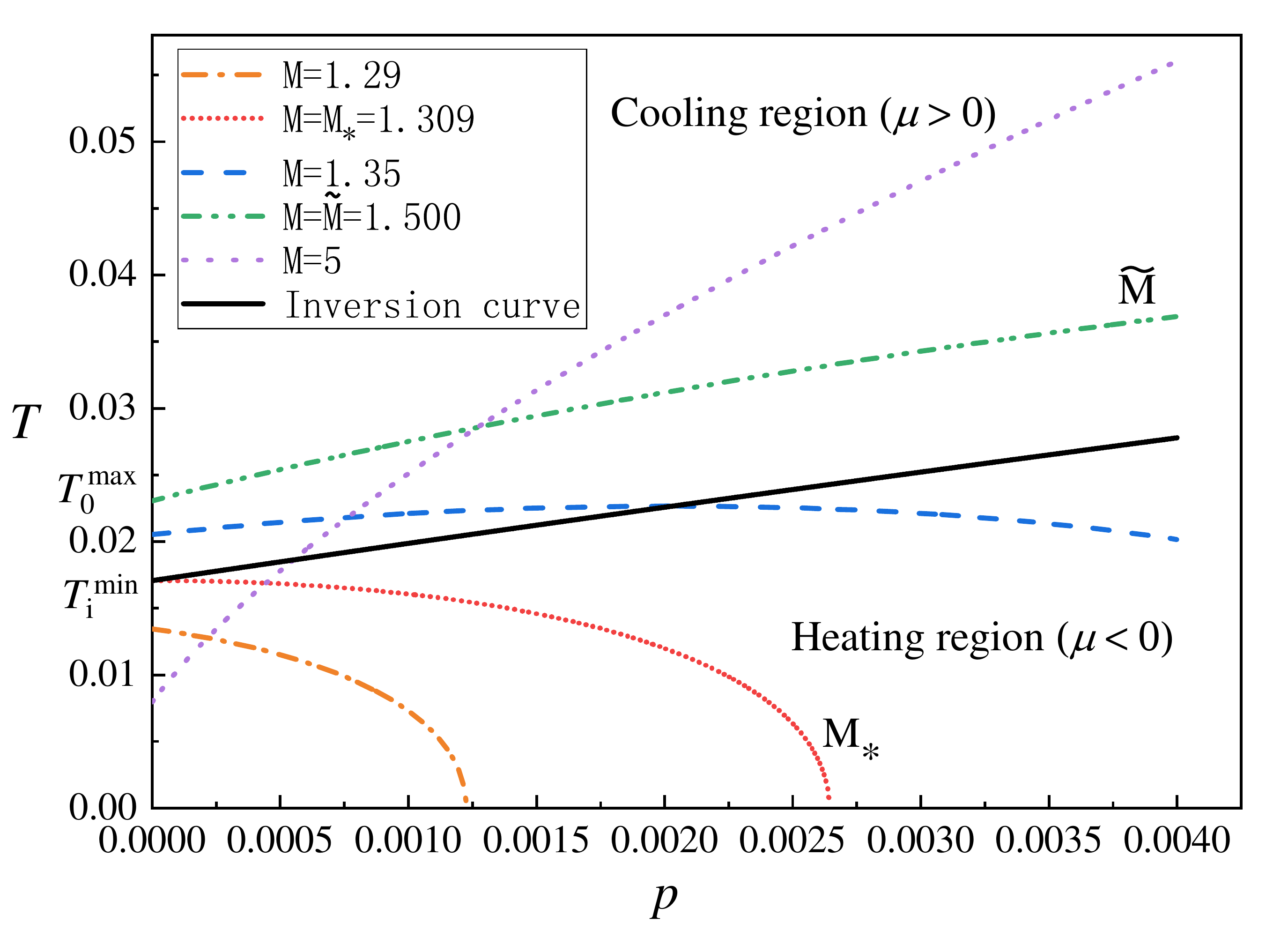}
\end{center}
\caption{The throttling process of the KN--AdS black holes with different enthalpies (i.e., masses), with $J=1$ and $Q=1$. The inversion curve is a monotonically increasing function and separates the $T$--$p$ plane into the cooling region ($\mu>0$) and the heating region ($\mu<0$). There is only a minimum inversion temperature $T_{\rm i}^{\rm min}=0.01707$, but no maximum one. Five characteristic isenthalpic curves are shown in order: (1) $M=1.29<M_\ast$, the slope of the isenthalpic curve is negative, and this curve always lies in the heating region; (2) $M=M_\ast=1.309$, the isenthalpic curve has the minimum inversion temperature $T_{\rm i}^{\rm min}$; (3) $M_\ast<M=1.35<\widetilde{M}$, the isenthalpic curve goes up and then down and crosses the inversion curve at its maximum point $(0.002028,0.02265)$; (4) $M=\widetilde{M}=1.500$, the isenthalpic curve has the largest $T$-intercept, $T_0^{\rm max}=0.02307$; (5) $M=5>\widetilde{M}$, the $T$-intercept decreases, and the isenthalpic curve intersects with other four isenthalpic curves. All isenthalpic curves with $M>M_\ast$ will descend and intersect with the $p$-axis at larger pressures, but cannot be shown in this figure.} \label{full}
\end{figure}

Finally, we give a complete description of the throttling process of the KN--AdS black holes in the extended phase space. In Fig. \ref{full}, the inversion curve and five characteristic isenthalpic curves are shown together by numerical methods (we choose $J=1$ and $Q=1$). The inversion curve is a monotonically increasing function, separating the $T$--$p$ plane into the cooling region with $\mu>0$ (above the inversion curve) and the heating region with $\mu<0$ (below the inversion curve). There is only a minimum inversion temperature $T_{\rm i}^{\rm min}$, but no maximum one. The isenthalpic curves are plotted with different enthalpies (i.e., masses) of the KN--AdS black holes: (1) $M<M_\ast$, the slope of the isenthalpic curve is negative, and this curve always lies in the heating region; (2) $M=M_\ast$, the isenthalpic curve possesses the minimum inversion temperature $T_{\rm i}^{\rm min}$; (3) $M_\ast<M<\widetilde{M}$, the isenthalpic curve goes up and then down and crosses the inversion curve at its maximum point; (4) $M=\widetilde{M}$, the isenthalpic curve has the largest $T$-intercept $T_0^{\rm max}$; (5) $M>\widetilde{M}$, the $T$-intercept is less than $T_0^{\rm max}$, and the isenthalpic curve intersects with other isenthalpic curves in the $T$--$p$ plane. In summary, the KN--AdS black holes exhibit highly rich and interesting behaviors in the throttling process depending on their masses. Last, we should also point out that the isenthalpic curves are relatively simple at high pressures, and the complexities above mainly appear at very low pressures. We believe this is the reason that they were not explored in detail before.

\section{Conclusion} \label{sec:con}

The black hole thermodynamics in the extended phase space has attracted increasing interest in recent years, in which the cosmological constant in the AdS space-time is interpreted as a dynamic variable and thus provides a varying thermodynamic pressure. In this framework, the equations of state of various black holes exhibit the non-ideal fluid characters, so the black holes may have abundant thermodynamic behaviors. The throttling process (i.e., JT effect) is a typical thermodynamic process, in which the enthalpies of the fluid are unchanged in the initial and final states, although the process is fundamentally irreversible. The black holes in the extended phase space may also experience the throttling process, and it should be simply understood as an adiabatic and isenthalpic process, in which the black hole enthalpies (i.e., masses) keep invariant.

In the present work, we systematically study the throttling process of the KN--AdS black holes---the most general rotating and charged black hole solution in four-dimensional space-time. The previous studies \cite{okcu1, D'Almeida, Gh, Mo, Chabab, Mo2018, Lan} mainly focused on the black hole solutions with simple spherical horizon topology. However, we first point out that the basic reason for these limitations is not physical but mathematical, and then apply two mathematical tricks to significantly simplify all the relevant calculations. By this means, the JT coefficient, inversion temperature, inversion curve, and isenthalpic curve of the KN--AdS black holes are carefully investigated in order, both analytically and numerically. We find that there are no maximum inversion temperatures, but only minimum ones that are around one half of the critical temperatures of the KN--AdS black holes, depending on the specific values of charges and angular momenta, but not very sensitively. Moreover, two notable and interesting issues not mentioned in the previous studies are also discussed in detail, namely the isenthalpic curves with negative slopes and the intersections of the isenthalpic curves, with two corresponding characteristic masses, $M_\ast$ and $\widetilde{M}$, calculated simultaneously. This implies that there are rich physical behaviors of black holes yet to be discovered. Altogether, we hope to present a complete picture and thorough understanding of the throttling process of the KN--AdS black holes. In addition, our calculation methods can be further applied to other various complicated black hole solutions, and we wish the present work would inspire more studies of black hole thermodynamics in the extended phase space.

Last, we should state that there still remains an important issue not explored in the present work---the phase transitions of the KN--AdS black holes in the throttling process. A preliminary study shows that this is a rather complicated problem. For example, the isenthalpic curves may even intersect with the coexistence lines between the small and large KN--AdS black holes more than once. This will be the topic of our future research.

\begin{acknowledgements}
We are very grateful to Zhao-Hui Chen, Tian-Fu Fu, Zhi-Zhen Wang, and Sheng-Jie Xuan for fruitful discussions. This work is supported by the Fundamental Research Funds for the Central Universities of China (No. N170504015).
\end{acknowledgements}

\end{document}